\documentclass[a4paper]{article}
\usepackage[a4paper, total={6.2in, 9in}]{geometry}
\usepackage{floatrow}
\usepackage{hyperref}
\usepackage{filecontents}
\usepackage{subfigure}
\usepackage{bm}
\usepackage[font={scriptsize,bf}]{caption}
\usepackage{amsmath}
\usepackage{amsfonts}
\usepackage{amssymb}
\usepackage{graphicx}
\usepackage{bm}
\usepackage{multicol}
\usepackage{multirow}
\usepackage{epstopdf}
\usepackage{soul}
\usepackage{pifont}
\usepackage{lineno}
\usepackage[table]{xcolor}
\usepackage{color}
\usepackage[autostyle]{csquotes}
\usepackage{tikz,pgf}
\usepackage{collcell}

\usepackage{pgfplots}
\usepackage{wasysym}
\usepackage{lipsum}
\usepackage{algorithm} 
\usepackage{algpseudocode} 
\newlength\fheight
\newlength\fwidth
\newcommand{\cmmnt}[1]{\ignorespaces}
\usepackage{etoolbox}
\usepackage{setspace}
\usepackage{csquotes}

\title{Neural Differential Equations for Oscillatory Flows in Aeroelasticity with Application to Transonic Buffet}
\author{ \large Michael Candon$^{1,2}$\thanks{corresponding author, candon.michael@rmit.edu.au}, Pier Marzocca$^1$ and Earl Dowell$^2$}

\date{
	\normalsize 
    $^1$Duke University, Durham, NC, 27708\\
    $^2$Department of Aerospace Engineering, RMIT University, Melbourne, AUS, 3000\\[2ex]%
}

\begin{document}
	
	\maketitle
	\begin{abstract}

        Self-excited aerodynamic flows arise across a broad range of systems and can drive nonlinear fluid–structure interactions and aeroelastic instabilities that are challenging and computationally expensive to predict. This paper presents a physics-guided neural differential equation (DE) reduced order model (ROM) combining a nonlinear fluid oscillator, a finite-memory multi-input Volterra series, and a compact neural network correction. The multi-input aerodynamic formulation is generalized to $m$ structural modes, capturing direct and nonlinear cross-modal coupling. The model is identified from a single prescribed-motion CFD simulation with simultaneous excitation of all retained structural modes, and is then coupled with the structural equations of motion for efficient aeroelastic prediction. Applied to transonic buffet over the ONERA OAT15A airfoil, the time-marching ROM predicts aeroelastic stability, frequency lock-in, and limit cycle amplitudes in good agreement with full-order reference solutions. The ROM is used to provide substantial new insight into buffet-induced aeroelastic instabilities involving more than one structural mode. 
        
    \end{abstract}

\section{Introduction}

Self-excited oscillatory flows occur across a wide range of fluid--structure systems, including vortex-induced vibration, cavitation, and transonic buffet. In these problems, autonomous flow oscillations interact with structural motion to produce nonlinear amplification, synchronization, and limit-cycle oscillations. Their prediction is challenging because the unsteady fluid loading depends on both intrinsic flow dynamics and structural dynamics that may also be history dependent. The discussion that follows is confined to transonic buffet, a representative and particularly demanding example in which these interactions are further complicated by unsteady transonic shock dynamics.

Transonic buffet is characterized by self-sustained shock oscillations and remains a major aircraft-design challenge because of its effects on structural fatigue and handling qualities~\cite{levinski20,candon22mssp}. Although buffet on rigid airfoils and wings has been studied extensively~\cite{giannelis17}, its interaction with elastic structures is less well understood. Naturally, buffet--structure interactions, whether under prescribed motion or elastic motion, are typically defined in terms of a frequency ratio $\hat{f}_n=\omega_n/\omega_B$, where $\omega_n$ is the imposed oscillation frequency or the natural frequency of structural mode $n$, and $\omega_B$ is the buffet frequency. Early forced-motion studies demonstrated frequency lock-in both above and below $\hat{f}_n=1$, where the shock oscillation synchronizes with the imposed structural motion~\cite{raveh09,hartmann13}. Subsequent numerical studies of two-dimensional elastically suspended airfoils showed that aeroelastic instability and lock-in can also occur in a single-degree-of-freedom pitching mode, but only for frequency ratios slightly above unity, typically persisting to $\hat{f}_\alpha\lesssim2$~\cite{raveh14,quan15,giannelis16}. Within this region, synchronization strongly amplifies the structural response, whereas outside it the response is generally much smaller and remains dominated by oscillations at the buffet frequency. This aeroelastic response asymmetry about $\hat{f}_\alpha = 1$ is important: while prescribed motion can enforce synchronization on either side of unity, the elastic system lock-in only occurs above $\hat{f}_\alpha = 1$, i.e., an aeroelastic instability must exist to drive the elastic response amplitude large enough to synchronization and lock-in. Gao~\textit{et al.}~\cite{gao17} interpreted this instability as a single-degree-of-freedom flutter mechanism, supported by Candon~\textit{et al.}~\cite{candon25f}, who demonstrated that negative aerodynamic damping drives the initial instability, with lock-in emerging as the unstable response grows. Buffet aeroelastic instabilities have also recently received experimental attention~\cite{themiot23,korthauer23}. 

Studies involving multiple structural degrees-of-freedom~\cite{raveh14, candon25g} remain limited, while fully three-dimensional investigations are rarer still~\cite{ionovich17}, largely because of the computational expense and numerical difficulty of resolving coupled buffet--structure dynamics. This motivates the development of efficient reduced-order models that can retain the essential nonlinear flow physics while enabling practical multi-mode aeroelastic analysis.

Reduced-order modeling of nonlinear unsteady aerodynamic and aeroelastic systems has evolved over several decades, with some prominent approaches including nonlinear oscillator models~\cite{hartlen70}, Volterra theory~\cite{silva97, silva05}, proper orthogonal decomposition~\cite{hall00}, dynamic mode decomposition~\cite{fonzi24}, harmonic balance~\cite{thomas04a}, and projection-based methods~\cite{carlberg13}. Despite this breadth, ROMs for transonic buffet remain scarce, particularly nonlinear models capable of predicting the full aeroelastic response. Buffet-only aerodynamic models based on nonlinear oscillators have been developed independently by Sansica \textit{et al.}~\cite{sansica22} and Ma \textit{et al.}~\cite{ma25}. The linearized aeroelastic model of Gao~\textit{et al.}~\cite{gao17} predicts stability but cannot reproduce limit-cycle oscillations. A practical nonlinear buffet aeroelastic ROM must simultaneously capture autonomous fluid oscillations, motion-induced coupling leading to flutter and lock-in, and nonlinear memory associated with shock motion and separation. Nonlinear oscillator models, long used for vortex-induced vibration~\cite{hartlen70,skop73,dowell81,marra11,hollenbach21}, naturally represent self-sustained flow oscillations but do not capture memory effects. Conversely, Volterra models provide an effective representation of nonlinear, history-dependent aerodynamic forces~\cite{silva97,balajewicz10,depaula19,candon24b,candon24c,candon25a}, but cannot independently generate autonomous flow oscillations. Candon~\textit{et al.}~\cite{candon25f,candon26a} combined these models in a Rayleigh--Volterra integro-differential model for transonic buffet aeroelasticity, with extensions based on sparse nonlinear system identification concepts of Brunton \textit{et al.}~\cite{brunton16}. The resulting formulation predicted aeroelastic stability and low-amplitude LCOs ($\lesssim 1^\circ$) with reasonable accuracy, but could not capture larger-amplitude responses and was restricted to a single structural degree of freedom. These limitations motivated the introduction of a neural correction.

In the original Neural ODE formulation of Chen~\textit{et al.}~\cite{chen18}, the governing vector field is represented by a neural network and the resulting trajectory is obtained through numerical integration. This idea was later extended to universal differential equations, in which neural networks augment known mechanistic models~\cite{rackauckas20}, and to neural controlled differential equations driven by external inputs~\cite{kidger20}, both of which are related to the physics-guided neural DE formulation used here.

This paper introduces a general-purpose physics-guided Neural DE framework for self-excited oscillatory flows in aeroelasticity. The model is extended to multiple structural modes and generalized aerodynamic force outputs, while also substantially improving its ability to predict strongly nonlinear responses. It combines a nonlinear fluid oscillator, a finite-memory Volterra representation of structural forcing, and a compact neural-network correction for the remaining dynamics. All parameters are identified jointly through backpropagation over discrete-time rollouts. To be explicit about the distinction from the original Neural ODE formulation of Chen, the neural network does not replace the full vector field, but instead augments the underlying Rayleigh--Volterra model. The finite-memory structural inputs also distinguish the present model from a conventional Neural ODE. The resulting multi-input model learns both direct and cross-modal aerodynamic coupling from a single prescribed-motion CFD simulation in which all retained structural modes are excited simultaneously. Once trained, the model is coupled with a general $m$-DOF structural system without requiring aeroelastic-response data for training. The framework is demonstrated for transonic buffet over the ONERA OAT15A airfoil and assessed through generalized-force prediction, aeroelastic stability, frequency lock-in, and LCO amplitude prediction. New insights into buffet induced aeroelastic instabilities are provided, specifically the single-degree-of-freedom stability of coupled heave-pitch mode shapes, and also towards instabilities with modal coupling.

\section{Multi-Input Neural Differential Equation Architecture}
\label{sec:DEROM}

For consistency with the CFD datasets and the identification procedure, all models in this section are written in discrete time. Time histories are sampled at uniform intervals $t_n=n\Delta t$ and $x_n := x(t_n)$. The quantities $\dot{x}_n$ and $\ddot{x}_n$ represent numerical estimates of the first and second time derivatives at $t_n$ (computed from the sampled data using finite differences).

\subsection{Model structure}
Let $\mathbf{\xi}_n=[\xi_{1,n},\ldots,\xi_{m,n}]^\mathsf{T}$ denote the $m$ structural coordinates, and let $Q_{i,n}$ denote the $i$th generalized aerodynamic force. At sample $n$, the dynamics of each aerodynamic force output are represented by a nonlinear dynamical system whose evolution depends on the current aerodynamic state, the current and past motion of all structural coordinates, and a data-driven correction:

\begin{equation}
\begin{aligned}
\dot{Q}_{i,n} &= V_{i,n} \\
\dot{V}_{i,n} &=
\mathcal{F}_{F,i}\!\left(Q_{i,n},V_{i,n}\right)
+
\mathcal{F}_{S,i}\!\left(\mathbf{z}_n\right)
+
\mathcal{F}_{\mathrm{NN},i}\!\left(Q_{i,n},V_{i,n},\mathbf{z}_n\right)
\end{aligned}
\label{eq:1}
\end{equation}

\noindent where

\begin{equation}
\mathbf{z}_n=
\left\{
\mathbf{\xi}_{n-k},
\dot{\mathbf{\xi}}_{n-k},
\ddot{\mathbf{\xi}}_{n-k}
\right\}_{k=0}^{N_L}
\label{eq:2}
\end{equation}

\noindent is the structural motion history vector containing the current and past motion of all $m$ structural coordinates. In a simple yet highly effective construction of such a system, $\mathcal{F}_F(\cdot)$ can be chosen as a canonical nonlinear oscillator to permit a self-sustained limit cycle, while a discrete-time pruned multi-input Volterra series $\mathcal{F}_S(\cdot)$ using series~\cite{balajewicz12}. can be represented using Parkinson's galloping model~\cite{parkinson89} 

The final term $\mathcal{F}_{\mathrm{NN}}(\cdot)$ is a small neural network that senses both the oscillator state and the full forcing vector and adds a global residual correction to the modeled acceleration, compensating for deficiencies in both the canonical Rayleigh oscillator structure and the chosen polynomial/lag forcing basis.

\subsection{Model Construction}

\subsubsection{Rayleigh Fluid Oscillator}

The intrinsic fluid dynamics are represented by a Rayleigh oscillator, which has been used extensively to model self-excited oscillatory flows in bluff-body aerodynamics~\cite{dowell81}. Accordingly, the flow-only contribution in Eq.~\eqref{eq:1} is defined as

\begin{equation}
\boxed{
\begin{aligned}
\dot{Q}_{i,n} &= V_{i,n},\\
\dot{V}_{i,n}
&=
\mathcal{F}_{F,i}\!\left(Q_{i,n},V_{i,n}\right)\\
&=
\epsilon_i\left(1-\alpha_iV_{i,n}^{2}\right)V_{i,n}
-\omega_F^2\left(Q_{i,n}-\bar{Q}_i\right)
\end{aligned}}
\label{eq:3}
\end{equation}

\noindent where

\begin{equation}
\alpha_i=\frac{1}{\omega_F^2Q_{i,\mathrm{ref}}^2},
\qquad
Q_{i,\mathrm{ref}}=Q_{i,\max}-\bar{Q}_i
\label{eq:4}
\end{equation}

\noindent and $\omega_F$ is the autonomous fluid-oscillation frequency, $\epsilon_i$ governs the linear growth and nonlinear saturation of generalized aerodynamic force output $i$, and $Q_{i,\mathrm{ref}}$ defines its reference oscillation amplitude about the mean. The negative linear damping produces growth from small disturbances, while the positive cubic damping saturates that growth to form a stable fluid limit cycle. For transonic buffet, $\omega_F$ corresponds to the shock oscillation frequency; for bluff-body aerodynamics, it corresponds to the wake frequency. Importantly, Eq.~\eqref{eq:3} constitutes a complete fluid-only ROM and may be used independently, without either structural dynamics or a neural network correction.

\subsubsection{Rayleigh-Volterra Fluid-Structural Oscillator}

The influence of the structural motion is represented using a discrete-time multi-input Volterra series. A complete $p$th-order multi-input Volterra operator contains $p$ independent lag coordinates and may be written as

\begin{equation}
\mathcal{V}_{i,p}^{\Xi}=
\sum_{j_1=1}^{m}\cdots\sum_{j_p=1}^{m}
\sum_{k_1=0}^{N_L}\cdots\sum_{k_p=0}^{N_L}
D_{i,p,\Xi}^{j_1\cdots j_p}
[k_1,\ldots,k_p]
\prod_{\ell=1}^{p}
\Xi_{j_\ell,n-k_\ell}
\label{eq:5}
\end{equation}

\noindent where $\Xi\in\{\xi,\dot{\xi}\}$ denotes either modal displacement or velocity, $i$ identifies the generalized aerodynamic force output, and $j_1,\ldots,j_p$ identify the structural input modes. The full kernel becomes increasingly expensive as its order, memory length, and number of structural modes increase. To obtain a tractable formulation, only the main diagonal of each multidimensional kernel is retained, such that $k_1=\cdots=k_p=k$. The resulting diagonally pruned operator is

\begin{equation}
\widetilde{\mathcal{V}}_{i,p}^{\Xi}=
\sum_{j_1=1}^{m}\cdots\sum_{j_p=1}^{m}
\sum_{k=0}^{N_L}
H_{i,p,\Xi}^{j_1\cdots j_p}[k]
\prod_{\ell=1}^{p}
\Xi_{j_\ell,n-k}
\label{eq:6}
\end{equation}

\noindent where $H_{i,p,\Xi}^{j_1\cdots j_p}[k]$ denotes the main diagonal of the corresponding full Volterra kernel. This pruning reduces the number of lag-dependent coefficients from order $(N_L+1)^p$ to order $(N_L+1)$ for each combination of inputs, while retaining nonlinear coupling between structural modes. Combining the Rayleigh oscillator with the pruned multi-input Volterra series gives the coupled fluid--structural dynamical system

\begin{equation}
\boxed{
\begin{aligned}
\dot{Q}_{i,n} &= V_{i,n},\\
\dot{V}_{i,n}
&=
\mathcal{F}_{F,i}\!\left(Q_{i,n},V_{i,n}\right)
+
\mathcal{F}_{S,i}\!\left(\mathbf{z}_n\right)\\
&=
\epsilon_i\left(1-\alpha_iV_{i,n}^{2}\right)V_{i,n}
-\omega_F^2\left(Q_{i,n}-\bar{Q}_i\right)
+
\sum_{j=1}^{m}A_{ij}\ddot{\xi}_{j,n} 
+
\sum_{\Xi\in\{\xi,\dot{\xi}\}}
\sum_{p=1}^{P}
\sum_{j_1=1}^{m}\cdots\sum_{j_p=1}^{m}
\sum_{k=0}^{N_L}
H_{i,p,\Xi}^{j_1\cdots j_p}[k]
\prod_{\ell=1}^{p}
\Xi_{j_\ell,n-k}
\end{aligned}}
\label{eq:7}
\end{equation}

\noindent where $A_{ij}$ describes the instantaneous acceleration contribution of structural mode $j$ to generalized aerodynamic force output $i$. Terms for which $j_1=\cdots=j_p$ are referred to as direct Volterra kernels. Terms containing two or more distinct modal indices are cross-kernels and describe nonlinear cross-modal aerodynamic coupling. The zeroth-order Volterra term is omitted because the static aerodynamic load is already represented by $\bar{Q}_i$. Equation~\eqref{eq:7} is itself a complete nonlinear fluid--structure ROM~\cite{candon25f}. A neural network correction is therefore not inherently required: when the Rayleigh--Volterra model provides sufficient accuracy, it should be used independently as a compact and interpretable aeroelastic ROM. The neural network introduced in the following section is therefore treated as an optional residual correction for dynamics that cannot be represented adequately by the selected oscillator and Volterra basis. 

It is also important to understand the relationship that this model has with the original models of this class~\cite{hartlen70, dowell81}. Specifically, considering a single structural mode ($m=1$), neglecting structural lags ($N_L=0$), retaining only odd-order structural terms ($p=1,3,5,\ldots$), and selecting velocity as the structural input ($\Xi=\dot{\xi}$), Eq.~\eqref{eq:7} reduces to a Rayleigh--Parkinson fluid--structure oscillator:

\begin{equation}
\boxed{
\begin{aligned}
\dot{Q}_n &= V_n,\\
\dot{V}_n
&=
\epsilon\left(1-\alpha V_n^{2}\right)V_n
-\omega_F^2\left(Q_n-\bar{Q}\right)
+A\ddot{\xi}_n
+B_1\dot{\xi}_n
+B_3\dot{\xi}_n^3
+B_5\dot{\xi}_n^5
+\hdots
B_p\dot{\xi}_n^p
\end{aligned}}
\label{eq:8}
\end{equation}

This reduced formulation is itself also a very useful and interpretable model for low-speed and other mildly nonlinear fluid--structure interactions, for which nonlinear memory effects are weak and a neural correction is unnecessary. Multi-input extensions follow naturally, as per Eqs.~\ref{eq:5}-\ref{eq:7}

\subsubsection{Rayleigh--Volterra Fluid--Structural Oscillator with neural network Correction}

The final contribution in Eq.~\eqref{eq:1} is a compact, fully connected feedforward neural network that corrects dynamics not represented by the Rayleigh oscillator or the pruned Volterra basis. For generalized aerodynamic force output $i$, the network input is formed from the instantaneous aerodynamic state and the complete structural motion history,

\begin{equation}
\mathbf{s}_{i,n}
=
\begin{bmatrix}
Q_{i,n} &
V_{i,n} &
\mathbf{z}_n^\mathsf{T}
\end{bmatrix}^{\mathsf T}
\label{eq:9}
\end{equation}

\noindent Two hidden layers with hyperbolic-tangent activation functions are followed by a linear scalar output layer

\begin{equation}
\mathbf{h}_{i,n}^{(1)}
=
\tanh\left(
\mathbf{W}_{i}^{(1)}\mathbf{s}_{i,n}
+
\mathbf{b}_{i}^{(1)}
\right)
\label{eq:10}
\end{equation}

\begin{equation}
\mathbf{h}_{i,n}^{(2)}
=
\tanh\left(
\mathbf{W}_{i}^{(2)}\mathbf{h}_{i,n}^{(1)}
+
\mathbf{b}_{i}^{(2)}
\right)
\label{eq:11}
\end{equation}

\begin{equation}
\mathcal{F}_{\mathrm{NN},i}
\left(Q_{i,n},V_{i,n},\mathbf{z}_n\right)
=
\mathbf{W}_{i}^{(3)}\mathbf{h}_{i,n}^{(2)}
+
b_{i}^{(3)}
\label{eq:12}
\end{equation}

\noindent The network therefore provides a residual correction to the modeled aerodynamic acceleration associated with generalized aerodynamic force output $i$. It is not trained separately against a prescribed residual; instead, its parameters are identified jointly with the Rayleigh and Volterra coefficients through backpropagation over the complete discrete-time rollout. $L_2$ regularization is applied to the neural network weights to improve generalization and encourage the network to remain a smooth correction to the physics-based backbone.

Combining the Rayleigh oscillator, multi-input Volterra series, and neural network correction gives

\begin{equation}
\boxed{
\begin{aligned}
\dot{Q}_{i,n} &= V_{i,n},\\
\dot{V}_{i,n}
&=
\mathcal{F}_{F,i}\!\left(Q_{i,n},V_{i,n}\right)
+
\mathcal{F}_{S,i}\!\left(\mathbf{z}_n\right)
+
\mathcal{F}_{\mathrm{NN},i}
\!\left(Q_{i,n},V_{i,n},\mathbf{z}_n\right)\\
&=
\epsilon_i\left(1-\alpha_iV_{i,n}^{2}\right)V_{i,n}
-\omega_F^2Q_{i,n}\\
&\quad+
\sum_{j=1}^{m}A_{ij}\ddot{\xi}_{j,n}
+
\sum_{\Xi\in\{\xi,\dot{\xi}\}}
\sum_{p=1}^{P}
\sum_{j_1=1}^{m}\cdots\sum_{j_p=1}^{m}
\sum_{k=0}^{N_L}
H_{i,p,\Xi}^{j_1\cdots j_p}[k]
\prod_{\ell=1}^{p}
\Xi_{j_\ell,n-k}\\
&\quad+
\mathcal{F}_{\mathrm{NN},i}
\left(Q_{i,n},V_{i,n},\mathbf{z}_n\right)
\end{aligned}}
\label{eq:13}
\end{equation}

\noindent The Rayleigh parameters $\omega_F$, $\epsilon_i$, and $\alpha_i$ may either be identified separately from flow-only data and held fixed or treated as trainable parameters and optimized jointly with the Volterra and neural network coefficients.

A constant contribution could, in principle, be included independently within each model component: through an equilibrium offset in the Rayleigh oscillator, a zeroth-order Volterra term, or the neural network output bias. Retaining multiple static terms introduces unnecessary parameter redundancy and reduces identifiability. The Rayleigh and Volterra static contributions are therefore set to zero, and the static aerodynamic load is absorbed exclusively by the neural network output bias $b_i^{(3)}$.

\subsubsection{Training}

Training data are generated from a single unsteady CFD simulation under prescribed multi-input structural motion. All retained structural modes are excited simultaneously using mutually orthogonal band-limited signals, allowing their individual and cross-modal aerodynamic contributions to be distinguished within a single dataset. All modal inputs are held at zero for the first $N_B$ samples so that the trajectory also contains several cycles of autonomous fluid oscillation before structural forcing is introduced. The generalized aerodynamic force histories are low-pass filtered to remove high-frequency numerical noise, and the required aerodynamic and structural derivatives are estimated using finite differences. The model is therefore identified entirely from prescribed-motion data, without requiring aeroelastic-response data or a separate CFD simulation for each structural mode.

The model is trained by minimizing the discrepancy between the predicted and reference aerodynamic trajectories over a discrete-time rollout. At each sample, Eq.~\eqref{eq:13} is advanced using a forward-Euler scheme. Defining the stacked aerodynamic state as

\begin{equation}
\mathbf{x}_n=
\begin{bmatrix}
\mathbf{Q}_n^\mathsf{T} &
\mathbf{V}_n^\mathsf{T}
\end{bmatrix}^{\mathsf T}
\label{eq:14}
\end{equation}

\noindent the training objective is the mean-squared trajectory error with an $L_2$ penalty applied to the neural network weights:

\begin{equation}
\mathcal{L}
=
\frac{1}{N}
\sum_{n=1}^{N}
\left\|
\hat{\mathbf{x}}_n-\mathbf{x}_n
\right\|_2^2
+
\lambda
\sum_i\sum_\ell
\left\|
\mathbf{W}_i^{(\ell)}
\right\|_F^2
\label{eq:15}
\end{equation}

\noindent The Rayleigh, Volterra, and neural network parameters are optimized jointly through backpropagation across the complete numerical rollout. Alternatively, the Rayleigh parameters may be identified independently from the unforced portion of the CFD trajectory and held fixed during training. Early stopping is applied using held-out validation data, with the parameters corresponding to the minimum validation error retained.

    \section{Computational Framework}
    
     The present study has been performed for the ONERA OAT15A airfoil, with experimental measurements available from the transonic wind tunnel of the Onera-Meudon Centre in France~\cite{jacquin09}. The experimental model is designed to study 2D buffet, with a chord length of $c = 0.23$m with a span of 0.78m and a thick trailing edge of 0.005$c$. Experiments have been performed over a Mach number range of $0.70 \leq M_\infty \leq 0.75$ over a wind-off angle-of-attack (AOA) sweep of $2.4^\circ \leq \alpha_0 \leq 3.91^\circ$ to determine the transonic buffet envelope onset at a Reynolds number of $Re_\infty = 3.0\times10^6$ (based on the chord length).

    \subsection{Computational Fluid Dynamics Model}
\begin{figure}[!h]
    	\centering
    	\includegraphics[width=0.5\textwidth]{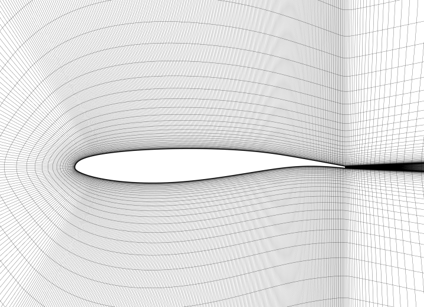}
    	\caption{Computational grid(47,400 cells)}
    	\label{fig:mesh}
    \end{figure}

    The general purpose finite volume code ANSYS Fluent 2024 R2~\cite{ansys} is used. The 2D URANS equations are solved using the density-based implicit solver with second-order upwind Roe-flux splitting scheme for the convective terms, and central-differencing for the diffusive terms. A dual time-stepping scheme is employed with second-order implicit temporal discretization and with a non-dimensional time-step of $\Delta \tau = 1\times 10^{-2}$ (100 steps per CTU). The computational grid (Fig.~\ref{fig:mesh}) is a structured C-grid topology with forced transition (represented by separate domains) imposed at ($x/c$) = 0.07. The average non-dimensional first cell height is $\bar{y^+} = 0.94$. The SST $k-\omega$ turbulence model is used with curvature correction~\cite{spalart97}. The convergence criteria is set to $1\times10^{-5}$ for the scaled residuals at each time-step. Validation, spatial and temporal refinement were conducted by Carrese \textit{et al.}~\cite{carrese16}.

\subsection{Aeroelastic Equations of Motion}

A two-dimensional typical-section model is used with heave, $h$, and pitch, $\alpha$, degrees of freedom. The structural equations are obtained from Lagrange's equations and written generally as

\begin{equation}
\mathbf{M}\ddot{\mathbf{x}}
+
\mathbf{C}\dot{\mathbf{x}}
+
\mathbf{K}\mathbf{x}
=
\mathbf{F}_a
\label{eq:16}
\end{equation}

\noindent where 

\begin{equation}
\mathbf{x}
=
\begin{bmatrix}
h\\
\alpha
\end{bmatrix},
\qquad
\mathbf{F}_a
=
\begin{bmatrix}
L\\
M_{c/4}
\end{bmatrix}
\label{eq:17}
\end{equation}

\noindent and the aerodynamic forces in physical coordinates are the lift, $L$, and moment about quarter chord, $M_{c/4}$. For the present typical section model, Eq.~\eqref{eq:16} becomes

\begin{equation} \begin{bmatrix} m & mx_\alpha b \\ mx_\alpha b & I_\alpha \end{bmatrix} \begin{bmatrix} \ddot{h} \\ \ddot{\alpha} \end{bmatrix} + \begin{bmatrix} c_{hh} & c_{h\alpha} \\ c_{h\alpha} & c_{\alpha\alpha} \end{bmatrix} \begin{bmatrix} \dot{h} \\ \dot{\alpha} \end{bmatrix} + \begin{bmatrix} k_h & 0 \\ 0 & k_\alpha \end{bmatrix} \begin{bmatrix} h \\ \alpha \end{bmatrix} = \begin{bmatrix} L \\ M_{c/4} \end{bmatrix} \label{eq:18} \end{equation}

\noindent Here, $m$ is the sectional mass, $I_\alpha$ is the sectional moment of inertia about the elastic axis, and $x_\alpha b$ is the offset between the center of mass and elastic axis. The quantities $\omega_h$ and $\omega_\alpha$ are the heave and pitch natural frequencies, while $\zeta_h$ and $\zeta_\alpha$ are the corresponding structural damping ratios. The structural-to-fluid mass ratio is $\mu=m/(\pi\rho b^2)=870$, based on the semi-chord $b=0.15$~m and freestream density $\rho=0.923$~kg/m$^3$. The elastic axis is located at $x/c=0.25$. These parameters are selected based on the work of Giannelis~\textit{et al.}~\cite{giannelis16}. The generalized aerodynamic forces associated with heave and pitch are denoted by $Q_h$ and $Q_\alpha$, respectively.

The physical coordinates are related to the modal coordinates through

\begin{equation}
\mathbf{x}
=
\boldsymbol{\Phi}\mathbf{\xi}
\label{eq:19}
\end{equation}

\noindent where $\boldsymbol{\Phi}$ contains the structural mode shapes and $\mathbf{\xi}$ contains the modal coordinates. Substitution into Eq.~\eqref{eq:16} and pre-multiplication by $\boldsymbol{\Phi}^{\mathsf T}$ give

\begin{equation}
\mathbf{M}_\xi\ddot{\mathbf{\xi}}
+
\mathbf{C}_\xi\dot{\mathbf{\xi}}
+
\mathbf{K}_\xi\mathbf{\xi}
=
\mathbf{Q}
\label{eq:20}
\end{equation}

\noindent where
$\mathbf{M}_\xi=\boldsymbol{\Phi}^{\mathsf T}\mathbf{M}\boldsymbol{\Phi}$,
$\mathbf{C}_\xi=\boldsymbol{\Phi}^{\mathsf T}\mathbf{C}\boldsymbol{\Phi}$,
$\mathbf{K}_\xi=\boldsymbol{\Phi}^{\mathsf T}\mathbf{K}\boldsymbol{\Phi}$, and $
\mathbf{Q} = \boldsymbol{\Phi}^{\mathsf T}\mathbf{F}_a$

Results are presented throughout in both physical and modal coordinates. The modal representation is important for determining whether an instability in the structurally coupled system is dominated by a single structural mode, or arises through two-mode coupling. For the full-order aeroelastic simulations, the structural equations of motion are embedded in ANSYS Fluent through a User-Defined Function and integrated using a fourth-order Runge--Kutta scheme with the aerodynamic loads frozen over each physical time step. Airfoil motion is accommodated using diffusion-based dynamic-mesh smoothing, which preserves mesh quality near the surface while dissipating the deformation toward the far field. The same structural time-integration scheme is used for the ROM-based aeroelastic simulations.

\clearpage
\section{Results}

The results focus on two challenging regimes for which the neural correction becomes important: (i) large-amplitude single-degree-of-freedom pitching response, and (ii) moderate-amplitude multi-degree-of-freedom aeroelastic response. All cases are evaluated at a freestream Mach number $M_\infty=0.73$, Reynolds number $Re_\infty\approx3\times10^6$ based on chord, and wind-off angle of attack $\alpha_0=3.5^\circ$. The baseline fluid-to-structural mass ratio is held at $\mu = 870$, unless otherwise specified.

\subsection{Single-Degree-of-Freedom Pitch System}

The single-degree-of-freedom pitching system is obtained from the two-degree-of-freedom typical-section model in Eq.~\eqref{eq:16} by constraining the heave coordinate, such that $h=\dot{h}=\ddot{h}=0$. A small structural damping ratio is used to permit the development of large-amplitude LCOs and provide a demanding assessment of the ROM. 

\subsubsection{Training and Cross-Validation}
Training is conducted using forced response data with band-limited random excitation. The frequency band chosen is $0.8 \leq \hat{f} \leq 2.0$ meaning the trained aerodynamic model is intended to be applied to aeroelastic models with natural frequencies in this same range. Both training and validation sets are used which are down-sampled by a factor of 10 from the CFD data to $\Delta \tau = 1\times10^{-1}$. In this work, when applied=, the NN correction employs 16 hidden units per layer. Similarly, when applied, the Volterra series includes 50 lags, corresponding to $\tau = 5$ (chosen based on the authors prior work~\cite{candon25f} where it was shown that optimal values were within the range $\tau \approx 2-10$. Numerous model variants are trained, all of which are optimized with back propagation, broadly fitting into five classes:

\begin{enumerate}
\item Rayleigh Oscillator with Parkinson's galloping model (without the NN)
\item Rayleigh Oscillator with Volterra series (without the NN)
\item Rayleigh Oscillator with NN (without the Volterra series)
\item Rayleigh Oscillator with Parkinson's galloping model and NN correction. 
\item Rayleigh Oscillator with Volterra series and NN correction. 
\end{enumerate}

\begin{figure}[!h]
    	\centering
    	\includegraphics[width=0.75\textwidth]{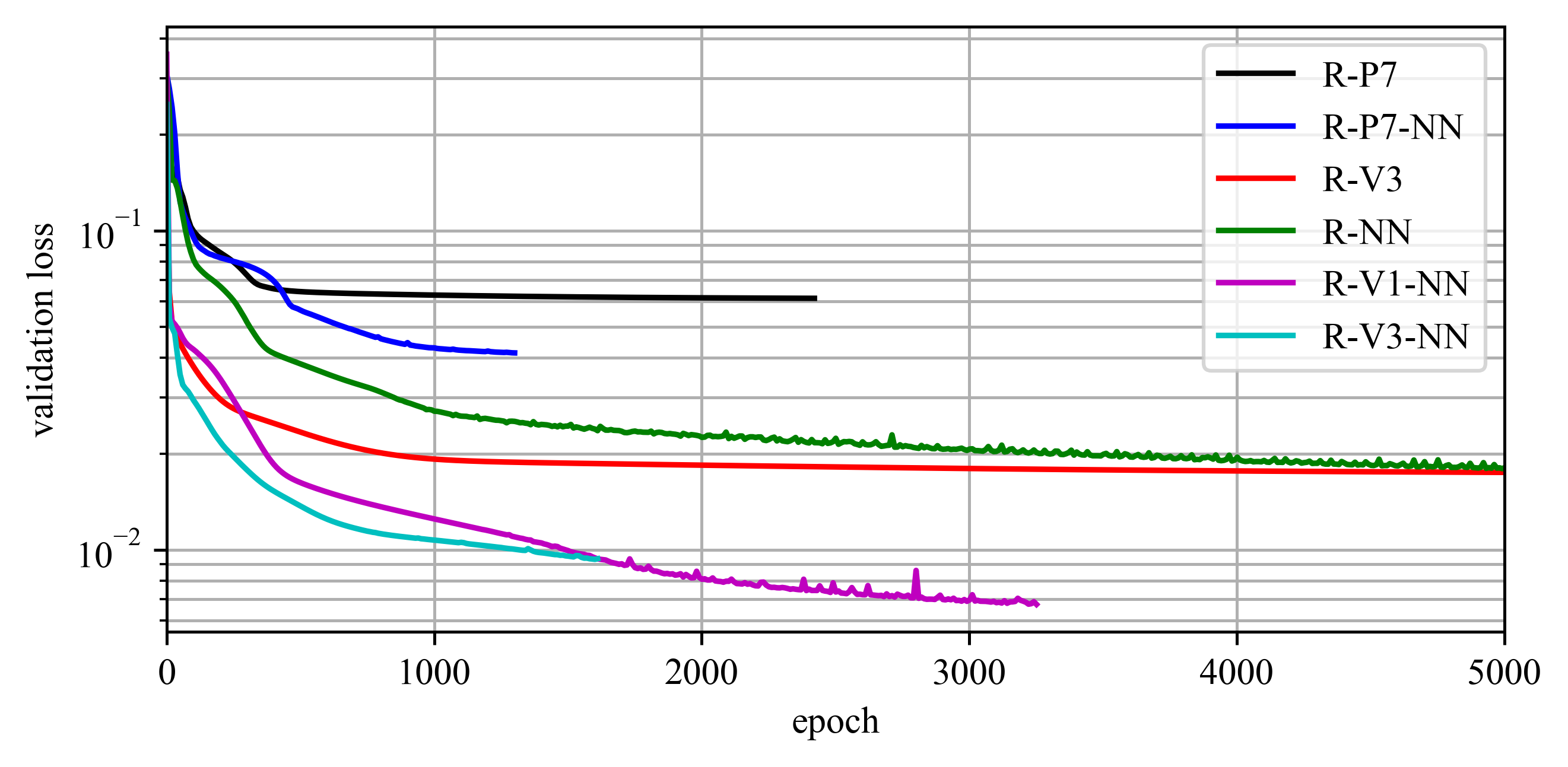}
    	\caption{Validation loss histories for the three model variants during training.}
    	\label{fig:loss}
    \end{figure}

The validation set loss for the different models is presented in Fig.~\ref{fig:loss} and the corresponding normalized root mean squared deviation (NRMSD) between the CFD and predicated validation datasets are given in Table~\ref{tab:nrmsd}. Based on the authors experience, it is recommended that the cross-validation NRMSD does not exceed 5\% in order to obtain a meaningful aeroelastic prediction. Ideally, it should be less than 3\%, and less than 2\% produced excellent results. 

The Rayleigh--Parkinson models perform worst and, while the neural correction does improve its performance substantially, it remains less accurate than the Volterra-based models. The weaker performance of the Rayleigh--Parkinson model for transonic buffet is already known~\cite{candon25f} and is unsurprising, as it is intended for very mildly nonlinear flows and does not capture transonic aerodynamic memory. The Rayleigh--Volterra (without the NN) and Rayleigh--NN (without the Volterra series) models achieve similar validation performance, although the Rayleigh--Volterra model converges more rapidly. Their NRMSD values are 5.04\% and 4.57\% respectively, lying near the upper limit for obtaining useful aeroelastic predictions. A key factor in the reasonable performance of the Rayleigh–Volterra model is identification of the coefficients through end-to-end backpropagation. When the same model is identified using a conventional least-squares fit, the validation error is substantially larger and the resulting model is not suitable for aeroelastic prediction. Among the Rayleigh--Volterra--NN models, R-V1-NN achieves the lowest NRMSD. Increasing the Volterra order to two or three does not improve validation accuracy, suggesting that the first-order memory terms capture the dominant history dependence while the neural correction accounts for the remaining nonlinear dynamics. The higher-order models therefore add complexity without a corresponding gain in predictive performance.

\begin{table}[!ht]
	\centering
	\caption{Validation NRMSD for the various trained model variants.}
    \label{tab:nrmsd}
	\begin{tabular}{clcccc}
		\hline
		Model & Description &  lags $N_L$& order $p$ & HU &NRMSD [\%]  \\
		\hline
        R-P7   & Rayleigh + Parkinson & 0 & 7 & -- & 10.09  \\
        R-V3   & Rayleigh + Volterra & 50 & 3 & -- & 5.04  \\
		R-NN   & Rayleigh + NN & -- & -- & 16 & 4.57  \\
        R-P7-NN   & Rayleigh + Parkinson + NN & 0 & 7 & 16 & 7.54  \\
        R-V1-NN   & Rayleigh + Volterra + NN & 50 & 1 & 16 & 2.75  \\
        R-V2-NN   & Rayleigh + Volterra + NN & 50 & 2 & 16 & 3.06  \\
        R-V3-NN   & Rayleigh + Volterra + NN & 50 & 3 & 16 & 3.30  \\
		\hline\hline		
	\end{tabular}
\end{table}

\begin{figure}[!h]
    \centering
    \includegraphics[width=1.0\textwidth]{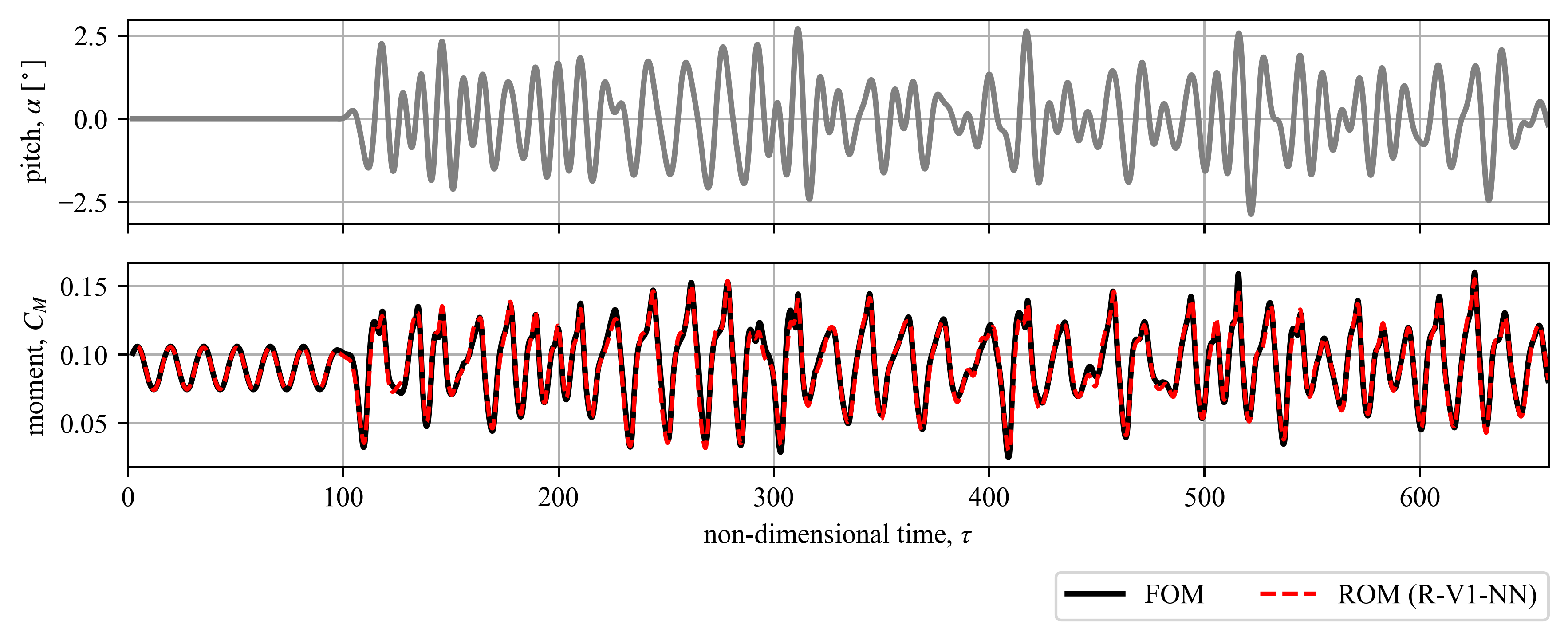}
    \caption{Time histories for a cross-validation case comparing the FOM (CFD) and R-V1-NN solutions.}
    \label{fig:crossval}
\end{figure}

Figure~\ref{fig:crossval} shows cross-validation dataset where the R-V1-NN (Rayleigh + 1st-order Volterra kernel + NN) solution closely matches the CFD-based full-order model (FOM) result. Given the highly pronounced nonlinear behavior visible even upon inspection of the time history, this result is very encouraging.

\subsubsection{Aeroelastic Modeling}

The trained unsteady aerodynamic ROM is now applied to an elastically suspended single-degree-of-freedom pitching system. The ability of the ROM to reproduce the lock-in region and the associated LCOs is assessed through a sweep of the structural natural frequency. The purpose of this section is to determine whether the ROM can predict the well-established large amplitude single-degree-of-freedom buffet aeroelastic behavior described in the introduction. 

\begin{figure}[!h]
    \centering
    \includegraphics[width=1.0\textwidth]{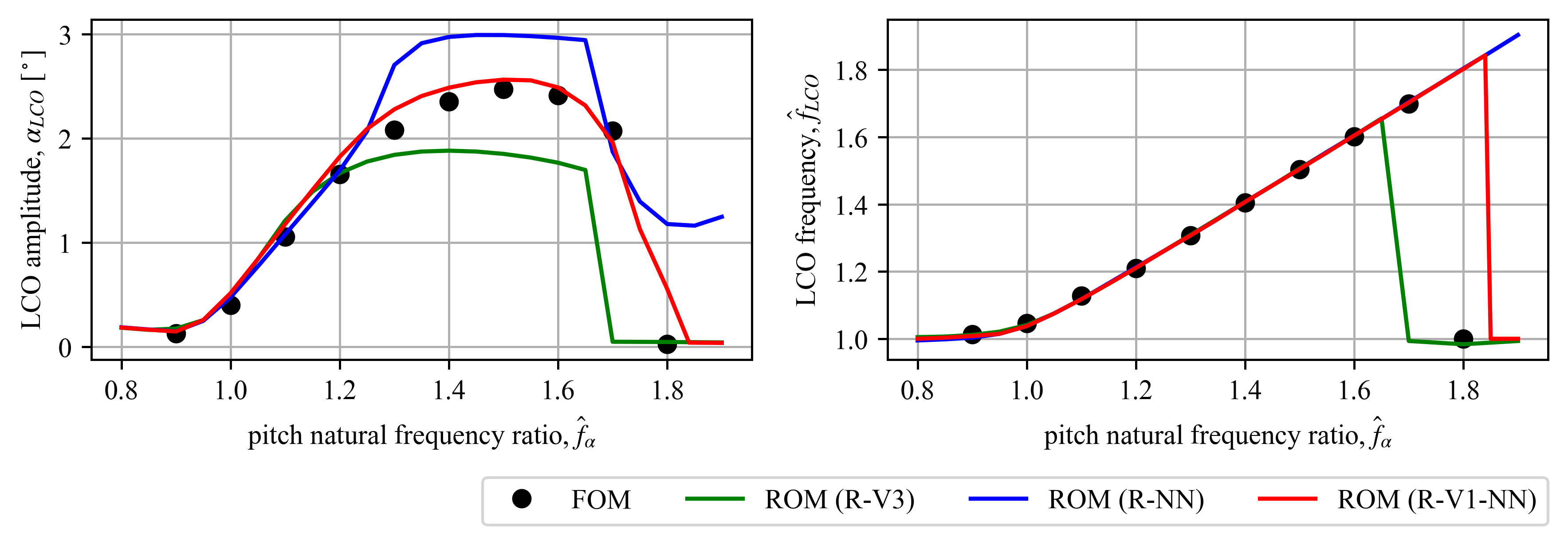}
    \caption{Comparison of FOM and ROM predictions of LCO amplitude and frequency versus natural frequency ratio with $\zeta_\alpha = 0.001$ for different ROM variants.}
    \label{fig:lcoa}
\end{figure}

Figure~\ref{fig:lcoa} compares the FOM predictions of LCO amplitude and frequency with those obtained from the three ROM variants. In terms of amplitude, the R-V1-NN model provides the best overall agreement with the FOM across the lock-in region, capturing both the onset of the instability and the peak response more accurately than the reduced variants. The Rayleigh + Volterra and Rayleigh + NN models both perform reasonably well for smaller LCO amplitudes ($\alpha_{LCO} \lesssim 2^\circ$). However, beyond this regime, the Rayleigh + Volterra model consistently underpredicts both the LCO amplitude and the upper boundary of the lock-in region. In contrast, the Rayleigh + NN model overpredicts the LCO amplitude and, more importantly, exhibits increasingly non-physical behavior at higher frequency ratios, predicting spurious locked-in responses beyond the boundary observed in the FOM. This suggests that, in the absence of the Volterra memory terms, the NN-only correction can fit parts of the response but does not preserve the correct qualitative aeroelastic behavior over the full parameter range. 

\begin{figure}[!h]
    \centering
    \includegraphics[width=1.0\textwidth]{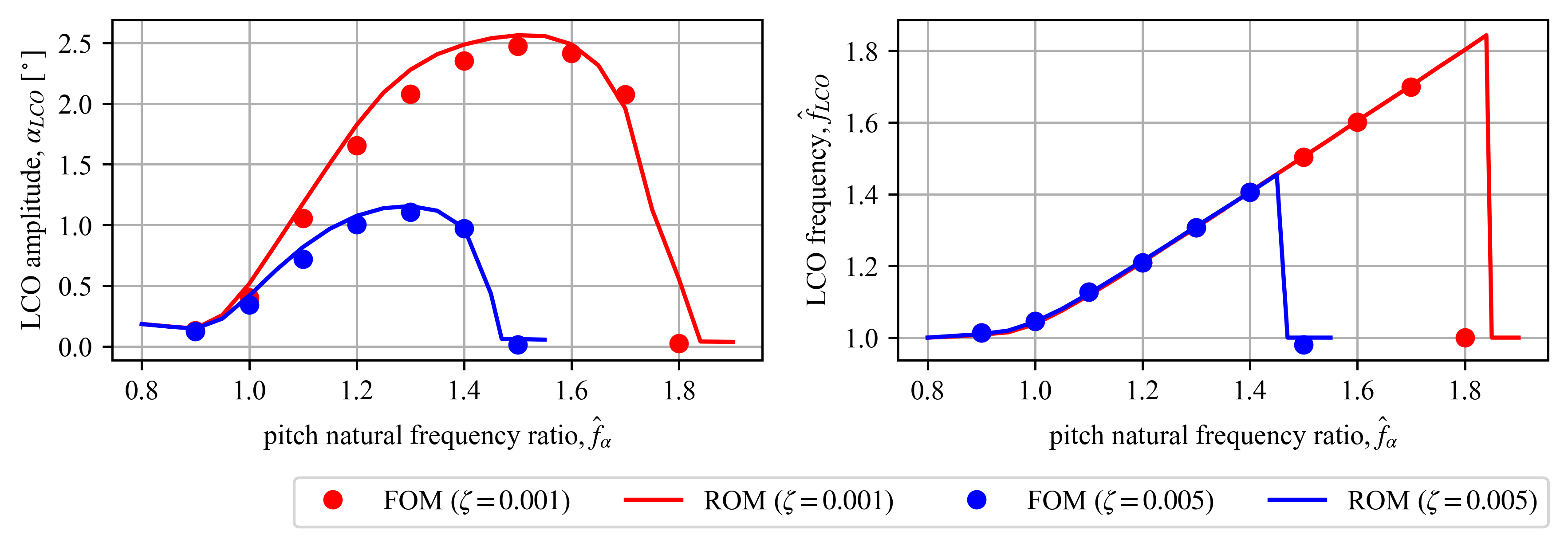}
    \caption{Comparison of FOM and ROM (R-V1-NN) predictions of LCO amplitude and frequency versus natural frequency ratio with different structural damping values.}
    \label{fig:lcob}
\end{figure}

Figure~\ref{fig:lcob} presents the predictions of the Rayleigh + Volterra + NN model with different structural damping ratios. The ROM accurately captures the expected stabilizing effect of increased structural damping, namely the reduction in LCO amplitude and the narrowing of the lock-in region, while also providing good agreement with the corresponding LCO frequencies. This is a particularly important result, since buffet-induced aeroelastic instability is governed by the balance between structural and aerodynamic contributions to the net effective damping, which is generally difficult to model accurately in a reduced order framework.

\begin{figure}[!h]
    \centering
    \includegraphics[width=1\textwidth]{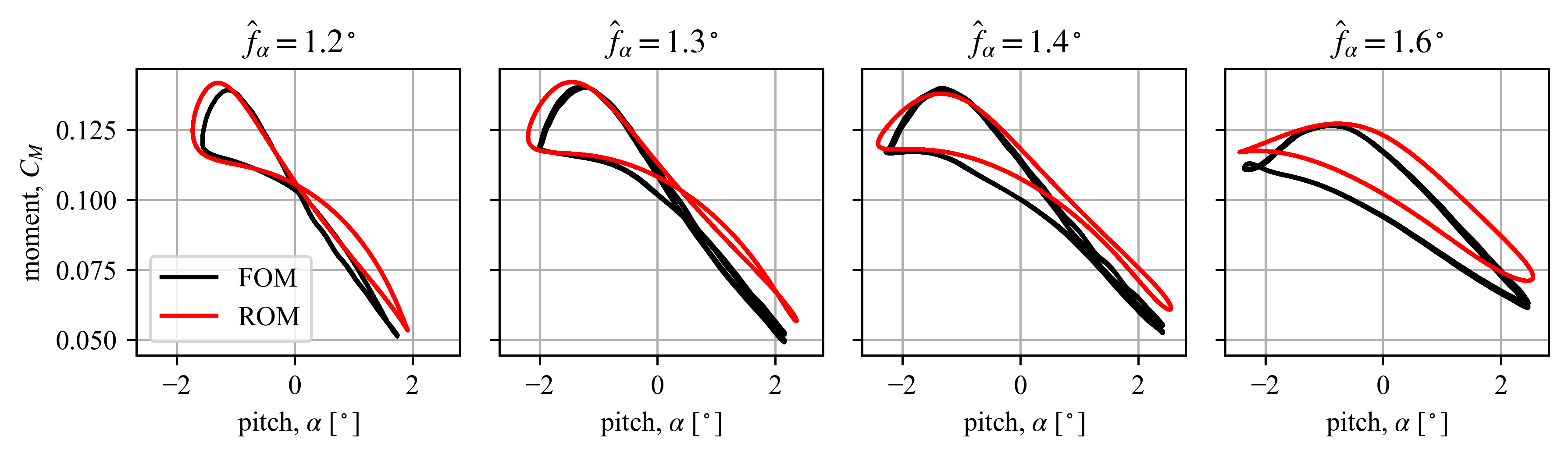}
    \caption{Lissajous curves of aerodynamic moment, $C_M$, versus pitch angle, $\alpha$, comparing the FOM and ROM (R-V1-NN) for representative natural frequency ratios.}
    \label{fig:liss}
\end{figure}

\begin{figure}[!h]
    \centering
    \includegraphics[width=1.0\textwidth]{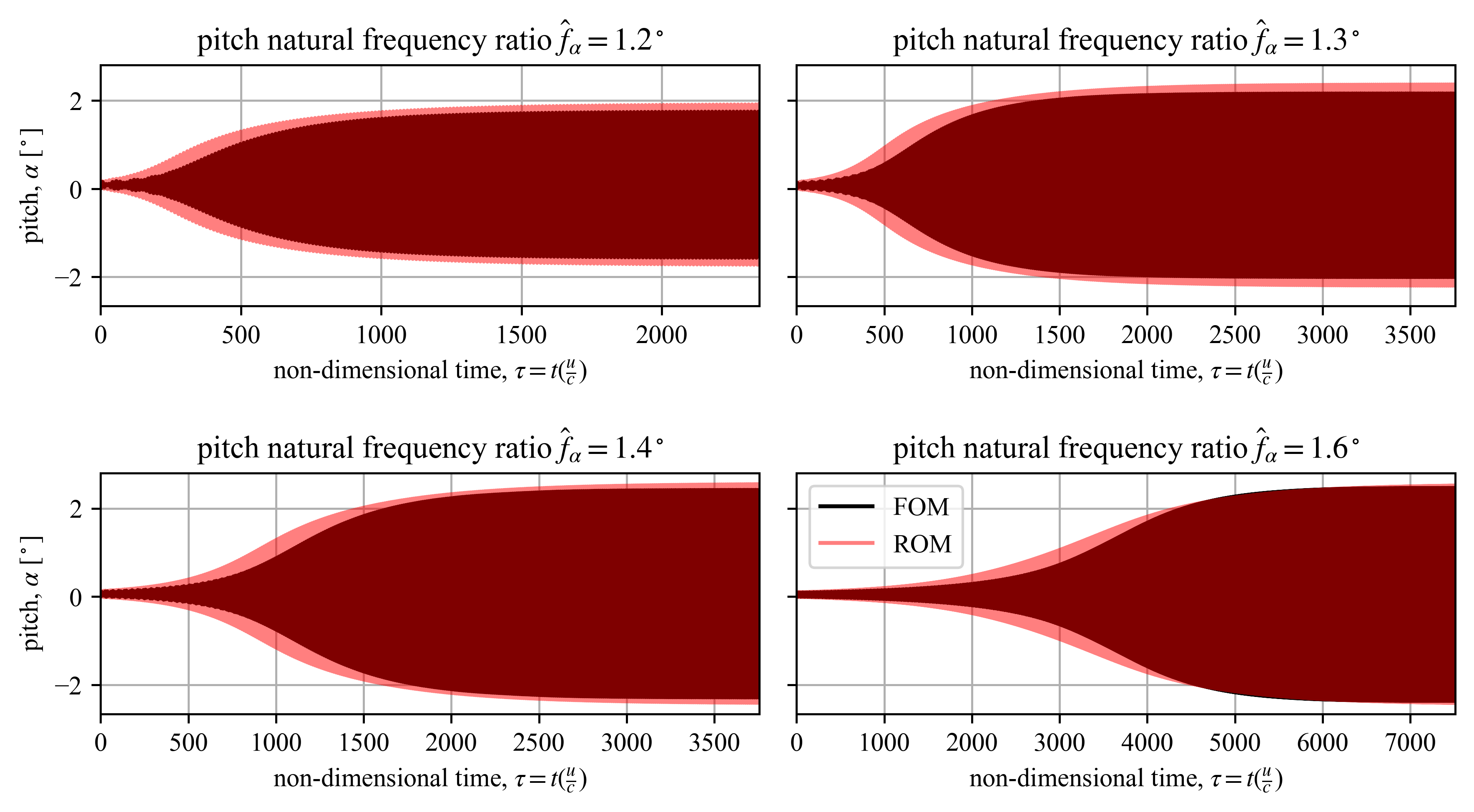}
    \caption{Comparison of FOM and ROM (R-V1-NN) predictions of pitch response time histories for representative natural frequency ratios.}
    \label{fig:time}
\end{figure}

Figures~\ref{fig:liss} and~\ref{fig:time} provide a more detailed comparison of the FOM and Rayleigh + Volterra + NN ROM predictions in both the time and phase domains. Overall, the ROM captures the transient growth toward the LCO and the final saturated response well, which is particularly encouraging because the transient is governed by the net effective damping and is therefore difficult to reproduce accurately in a reduced order model. The Lissajous curves further show that the ROM recovers the main nonlinear features of the aerodynamic moment response, including the phase lag, hysteresis, and overall loop shape. Although some discrepancies remain in the detailed loop geometry and in the exact amplitude evolution, this is not unexpected given the complexity of the underlying transonic flow physics, including strongly nonlinear shock motion, separation effects, and finite-memory aerodynamic behavior. In that context, the overall level of agreement is considered very good.

\clearpage
\subsection{Two-Degree-of-Freedom Heave--Pitch System}

The full two-degree-of-freedom typical-section model in Eq.~\eqref{eq:16} is now considered, with both heave and pitch free to respond. The analysis first examines single-degree-of-freedom instabilities, including cases in which inertial coupling produces mixed physical motion while the response remains dominated by a single structural mode. Coupled-mode instabilities are then examined. This section both verifies the multi-input ROM and provides new insight into coupled-mode buffet aeroelasticity.

\subsubsection{Training and Cross-Validation}
The prescribed-motion training data are generated in physical heave and pitch coordinates. This allows the same identified aerodynamic model to be projected onto any selected two-mode heave--pitch basis without repeating the CFD simulation. The excitation covers the normalized frequency range $0.3\leq\hat{f}\leq1.5$, with maximum amplitudes $h/b=0.05$ and $\alpha=0.0349$~rad $(2^\circ)$. Both physical coordinates are excited simultaneously using orthogonal band-limited signals.

Training the multi-input model is substantially more demanding because the input space must contain sufficient information to identify both direct and nonlinear cross-aerodynamic interactions. The training data was increased from a non-dimensional duration of $\tau\approx1200$ to $\tau\approx1800$. Because the single-degree-of-freedom study had already established the relative performance of the different model architectures, the full comparison of models with and without Volterra terms and neural corrections is not repeated. Despite the large number of possible nonlinear direct and cross-kernels in the multi-input Volterra expansion, the best performance was consistently obtained using only first-order kernels in heave velocity, $\dot{h}$, and pitch, $\alpha$. No explicit Volterra cross-kernels or higher-order kernels were required. Instead, the neural network accounted for the nonlinear and cross-modal contributions not represented by this compact linear-memory basis. 

\begin{figure}[h!]
    \centering
    \includegraphics[width=0.98\textwidth]{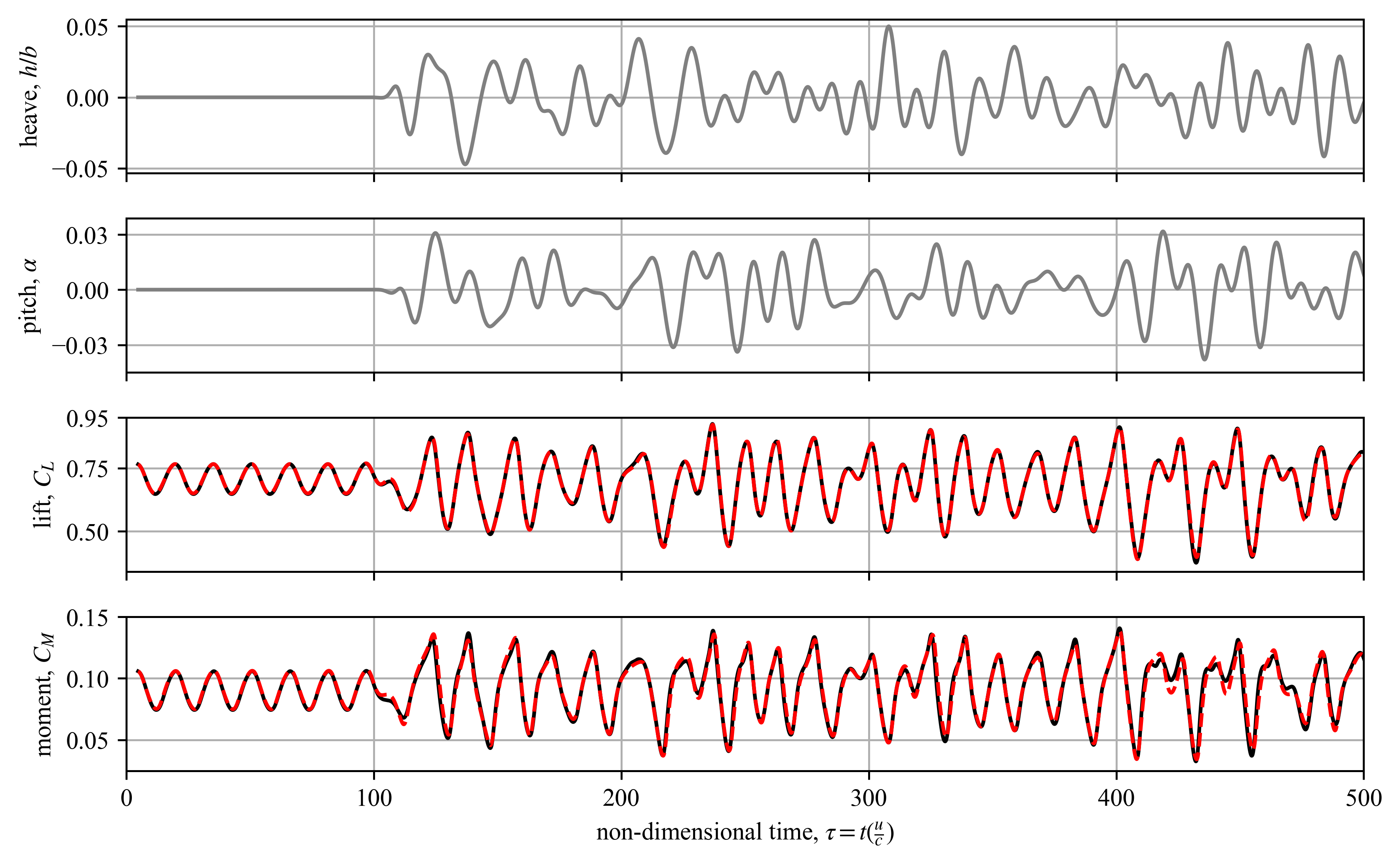}
    \caption{Cross-validation of the multi-input ROM under simultaneous prescribed heave and pitch excitation, comparing the CFD and ROM lift and moment responses}
    \label{fig:2dof_crossval}
\end{figure}

Figure~\ref{fig:2dof_crossval} presents a representative cross-validation trajectory, for which the multi-input ROM closely reproduces the CFD-based lift and moment responses under simultaneous excitation of both structural modes. The agreement in amplitude, phase, and temporal modulation demonstrates that the compact Volterra basis and neural correction together capture the principal multi-input aerodynamic dynamics.

\clearpage
\subsubsection{Single-Degree-of-Freedom Instabilities of Pure Heave and Pure Pitch Modes}

To isolate the pure heave/pitch mechanisms, the static unbalance is set to $x_\alpha=0$. These cases are intended to verify that the multi-input ROM can recover basic single-degree-of-freedom instabilities of a two-degree-of-freedom system, and to assess whether the presence of a second, stable structural mode alters the instability regime.

\begin{figure}[!h]
    \centering
    \includegraphics[width=1.0\textwidth]{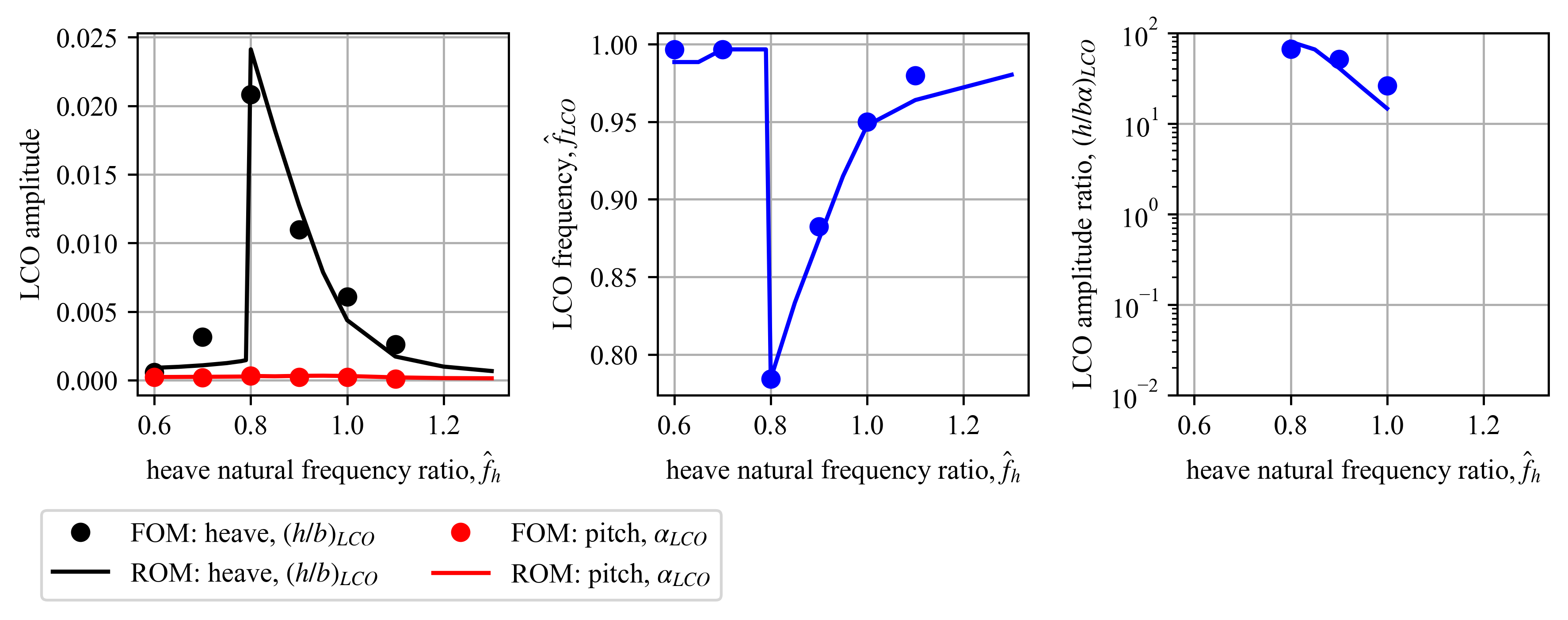}
    \caption{Pure-heave instability with $\hat{f}_\alpha=1.5$, $x_\alpha=0$, and $\zeta_h=\zeta_\alpha=0.005$, showing the LCO amplitudes, frequency, and heave-to-pitch amplitude ratio.}
    \label{fig:2dof_case1}
\end{figure}

The heave instability is considered first as presented in Fig.~\ref{fig:2dof_case1}. The pitch natural frequency is fixed at $\hat{f}_\alpha=1.5$ and, by applying structural damping of $\zeta_h=\zeta_\alpha=0.005$, the pitch mode remains stable at the baseline structural-to-fluid mass ratio. The heave natural frequency is swept through $\hat{f}_h=1$. As expected, the pure heave instability occurs below unity, consistent with previous single-degree-of-freedom results~\cite{candon25f}, and is essentially unaffected by the presence of the stable pitch mode. Within the instability region, $0.8 \leq \hat{f}_h \leq 1$, there is significant amplification of the heave mode which responds at $\hat{f}_h$. The heave amplitude exceeds the pitch response by approximately one to two orders of magnitude, as seen in the heave-to-pitch ratio plot, confirming that the instability is dominated by the heave mode. The ROM accurately reproduces the instability region together with the LCO amplitude and frequency.

\begin{figure}[!h]
    \centering
    \includegraphics[width=1.0\textwidth]{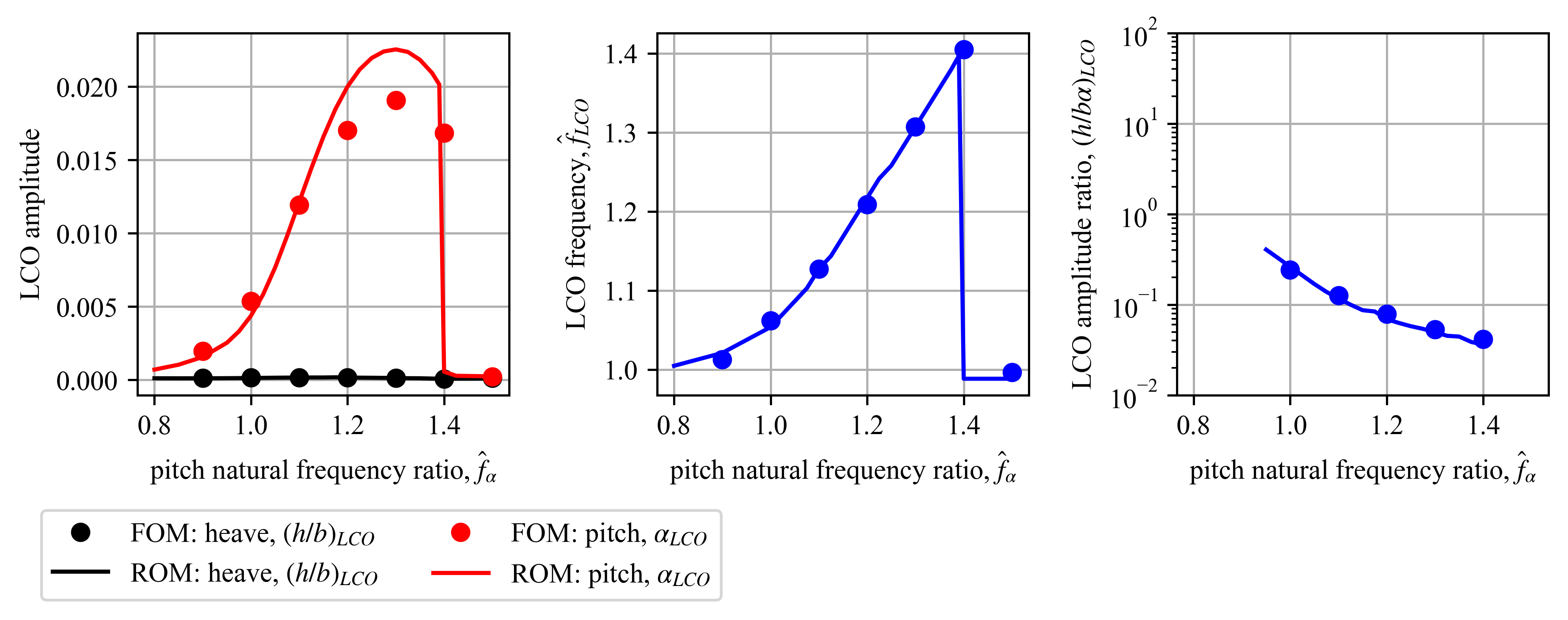}
    \caption{Pure-pitch instability with $\hat{f}_h=0.6$, $x_\alpha=0$, and $\zeta_h=\zeta_\alpha=0.005$, showing the LCO amplitudes, frequency, and heave-to-pitch amplitude ratio.}
    \label{fig:2dof_case2}
\end{figure}

The pure-pitch instability is examined next by fixing the heave natural frequency at $\hat{f}_h=0.6$ and sweeping the pitch natural frequency, with the same structural damping ratios $\zeta_h=\zeta_\alpha=0.005$ as is presented in Fig.~\ref{fig:2dof_case2}. The instability occurs above unity and persists to approximately $\hat{f}_\alpha=1.4$, consistent with the established single-degree-of-freedom pitch response. The heave-to-pitch amplitude ratio remains of order $10^{-1}$ and approaches $10^{-2}$ toward the upper boundary, confirming that the response is overwhelmingly pitch dominated. The instability region is therefore largely unaffected by the presence of the stable heave mode. The ROM captures the instability region, LCO amplitude and frequency reasonably well, although the amplitude prediction deteriorates somewhat near the upper boundary. Given the greater identification difficulty of the multi-input model, the overall agreement remains acceptable.

\clearpage

\subsubsection{Single-Degree-of-Freedom Instabilities of Coupled Heave-Pitch Modes}

Next, the center of gravity is shifted from $x_{CG}=x_{EA}=0.25$ to $x_{CG}=0.30$, corresponding to a static unbalance of $x_\alpha=0.43$. This introduces inertial coupling between the heave and pitch degrees-of-freedom in the physical structural equations. Although the linear structural modes remain uncoupled after transformation to modal coordinates, each corresponding mode shape contains both heave and pitch components. For the present system, mode 1 is heave dominated, whereas mode 2 is pitch dominated. While this configuration allows the stability of mixed heave--pitch mode shapes to be studied, as well as aerodynamic coupling between the modes, it does not guarantee that instabilities involves coupling between the modes.

\begin{figure}[!h]
    \centering
    \includegraphics[width=1.0\textwidth]{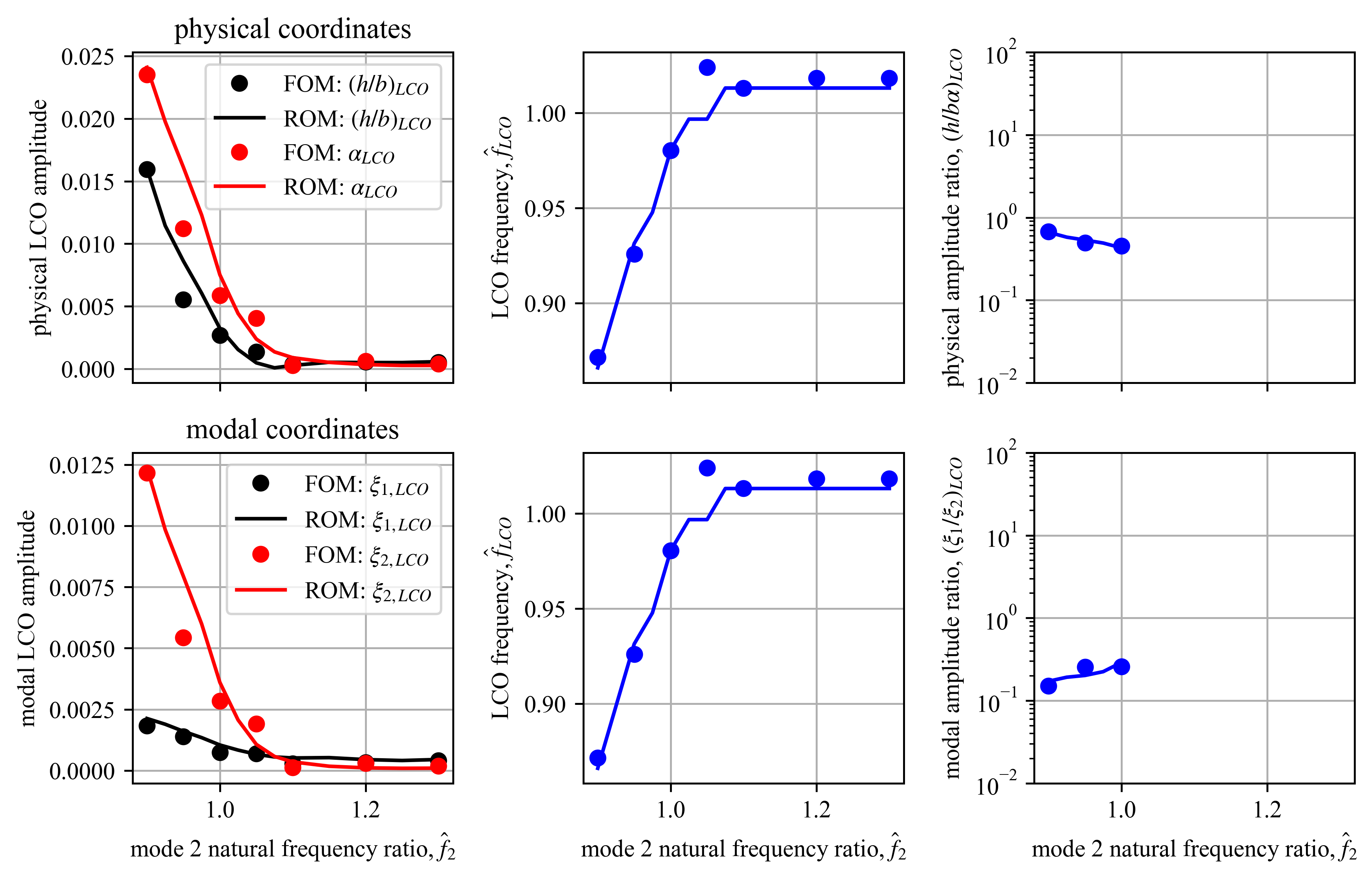}
    \caption{Mode 2 instability with $\hat{f}_1=0.515$, $x_\alpha=0.43$, and $\zeta_1=\zeta_2=0.00$, showing the LCO amplitudes, frequency, and amplitude ratios.}
    \label{fig:2dof_case3}
\end{figure}

Figure~\ref{fig:2dof_case3} shows that the ROM reproduces the FOM LCO trends in both the physical and modal coordinates. Although the physical response contains heave and pitch motion contributions, in modal coordinates, the response is strongly dominated by mode 2 and can be interpreted as a predominantly single-degree-of-freedom instability of the pitch-dominated second mode. The most important result is that this mode becomes unstable (experiences positive net aerodynamic work) for $\hat{f}_2<1$, which is the opposite of the behavior observed for a pure-pitch mode. This change is physically plausible because the net aerodynamic work for a mixed mode contains contributions from both the heave--lift and pitch--moment interactions. In the pure-mode cases, the heave--lift interaction produces positive work below unity, whereas the pitch--moment interaction produces positive work above unity. Therefore, for a mixed heave--pitch mode, the balance between these contributions can produce instability on either side of $\hat{f}_2=1$, even when the mode remains pitch dominated. The modal results also show modest amplification of mode 1, with $\xi_{1,LCO}/\xi_{2,LCO}$ of approximately $0.15$--$0.25$. Such participation was not observed for the pure-heave or pure-pitch configurations, indicating a degree of aerodynamic interaction between the modes, although mode 2 is clearly dominant.

\begin{figure}[!h]
    \centering
    \includegraphics[width=0.50\textwidth]{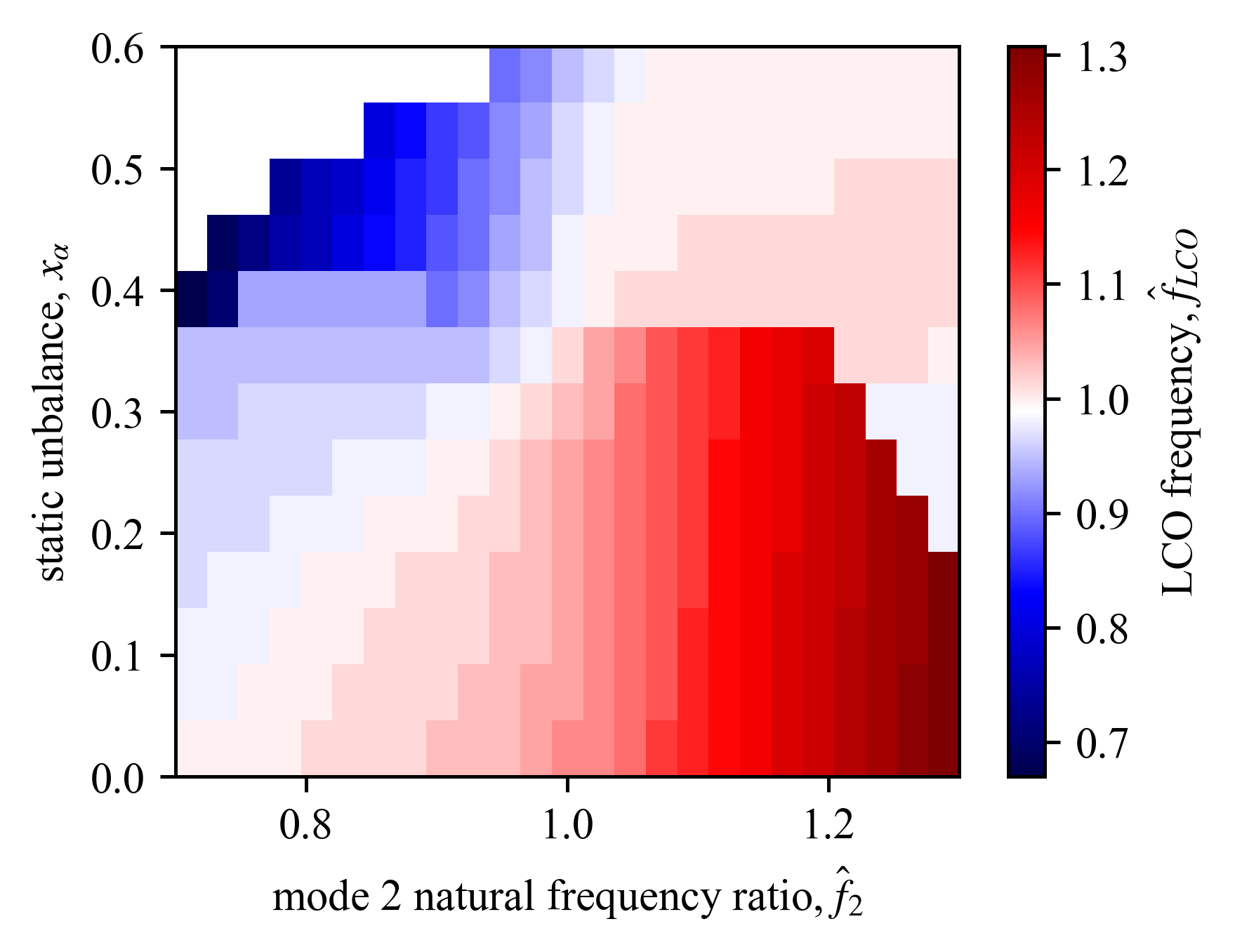}
    \caption{LCO frequency ratio as a function of the mode 2 natural frequency and static unbalance with $\hat{f}_1=0.4$ and $\zeta_1=\zeta_2=0.00$.}
    \label{fig:2dof_heatmap_frequency}
\end{figure}

\begin{figure}[!h]
    \centering
    \subfigure[Physical coordinates]{\label{phys}\includegraphics[width=0.80\textwidth]{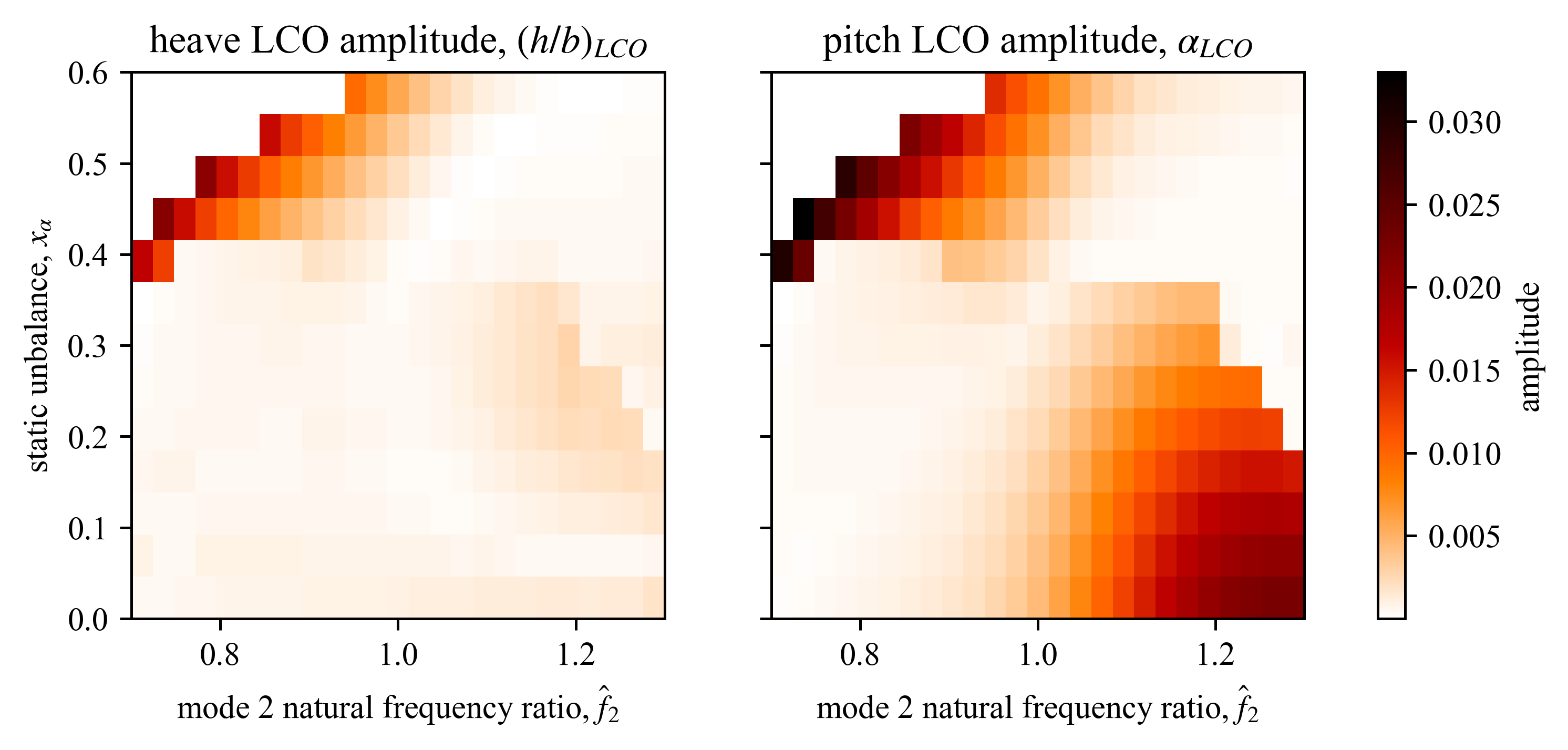}}

    \subfigure[Modal coordinates]{\label{modal}\includegraphics[width=0.80\textwidth]{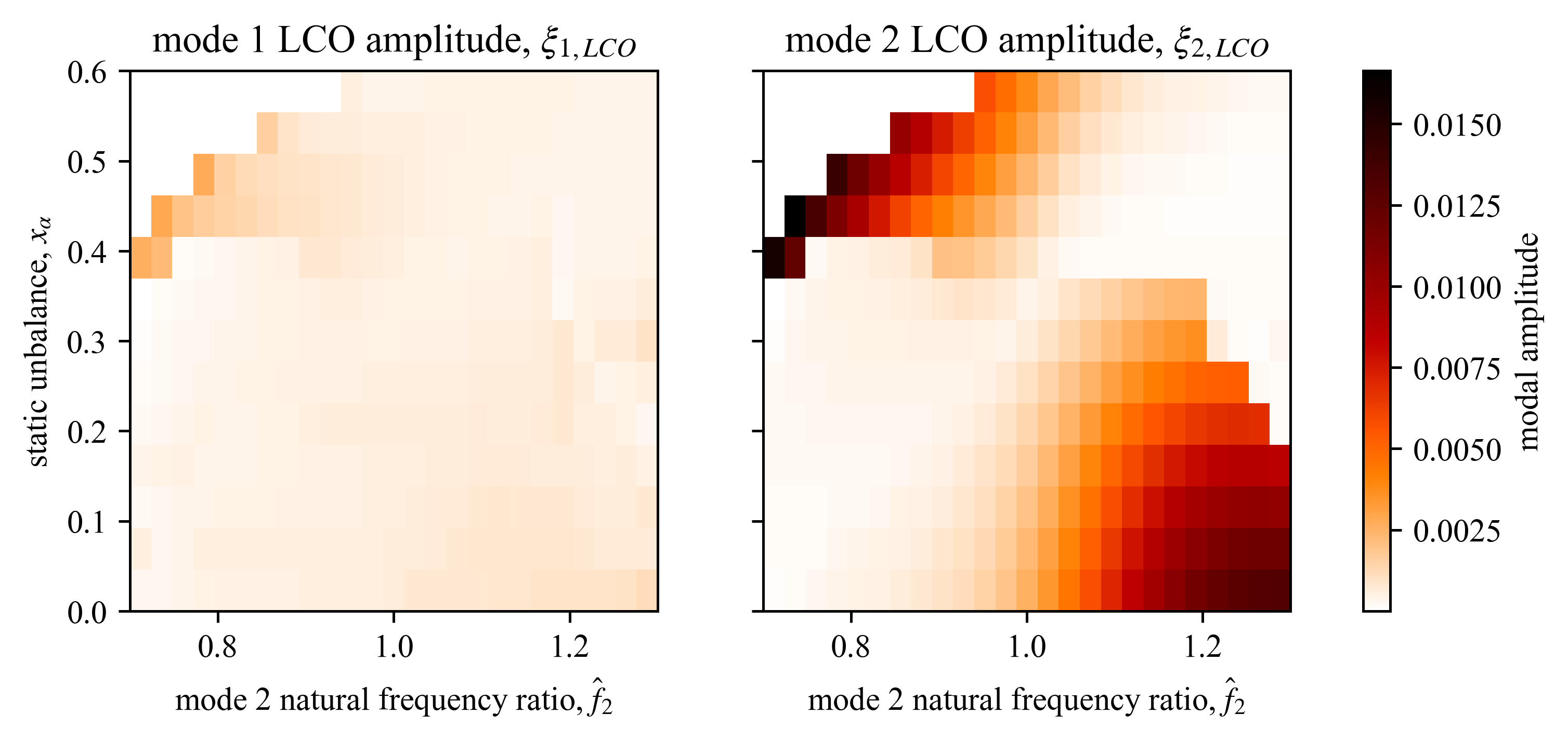}}
    \caption{LCO amplitudes as functions of the mode 2 natural frequency and static unbalance in the (a) physical and (b) modal coordinates with $\hat{f}_1=0.4$ and $\zeta_1=\zeta_2=0.00$.}
    \label{fig:2dof_heatmap_lcoamp}
\end{figure}

One of the main benefits of the ROM is that large-scale parametric studies can be performed very efficiently. Figures~\ref{fig:2dof_heatmap_frequency} and~\ref{fig:2dof_heatmap_lcoamp} present response maps of the LCO frequency and amplitude across the static unbalance and mode 2 natural frequency parameter space. The trend is clear: increasing the static unbalance shifts the unstable response of mode 2. At small $x_\alpha$, for which the structural modes remain close to pure heave and pure pitch, the mode 2 instability is concentrated above $\hat{f}_2=1$, which is consistent with the pure-pitch case. As $x_\alpha$ increases and mode 2 acquires a larger heave component, the instability shifts to $\hat{f}_2<1$. The change in the lock-in region, identified by $\hat{f}_{LCO}\approx1$, occurs between $x_\alpha=0.35$ and $0.40$. The modal coordinate results show that the instability is strongly dominated by mode 2 for the entire region considered here. Mode 1 remains weak over most of the parameter space, although a localized increase in its amplitude is observed for larger values of $x_\alpha$ and low values of $\hat{f}_2$, i.e., where the two structural natural frequencies are closest. These regions indicate increased aerodynamic interaction between the modes, but mode 2 remains the dominant contributor to the LCO. The principal conclusion from this part of the work, at least at this stage, is that single-degree-of-freedom instabilities are likely to dominate buffet-driven aeroelastic behavior. This is not surprising, considering that such instabilities are predominantly driven by negative aerodynamic damping and therefore can persist down to a dynamic pressure of zero in the absence of structural damping and flow viscosity.

\clearpage 

\subsubsection{Towards Coupled-Mode Instabilities}
Finally, findings on \underline{\textbf{potential}} coupled-mode aeroelastic instabilities arising in buffeting flows are presented. Considerable difficulty was encountered in identifying reasonable structural parameters for which both structural modes participated strongly. In most cases, one mode clearly dominated the response. It was even more difficult to identify such a case where the response amplitudes and frequencies remained within the range over which the ROM had been trained and validated. 

Initially, for the structural model, $x_\alpha = 0.15$ is chosen with mode 2 natural frequencies spanning $\hat{f}_2 = 0.9$, mode 1-to-2 natural frequency ratios spanning $\omega_1\, / \, \omega_2 = 0.79 - 0.81$, and no structural damping. Within this parameter space the aeroelastic system can exhibit stability\footnote{the definition of stable response here is that, without structural damping, the system responds at the buffet frequency with very low amplitude, and it is insensitive to any initial perturbation} at the nominal fluid-to-structural mass ratio, $1/\mu$, for which the model was trained. The analysis in this sections conducts a sweep of the fluid-to-structural mass ratio to study how the response developed. The fluid-to-structural mass ratio scales proportionally to freestream dynamic pressure, $1/\mu \propto q_\infty$.

\begin{figure}[h!]
    \centering
    \includegraphics[width=1.0\textwidth]{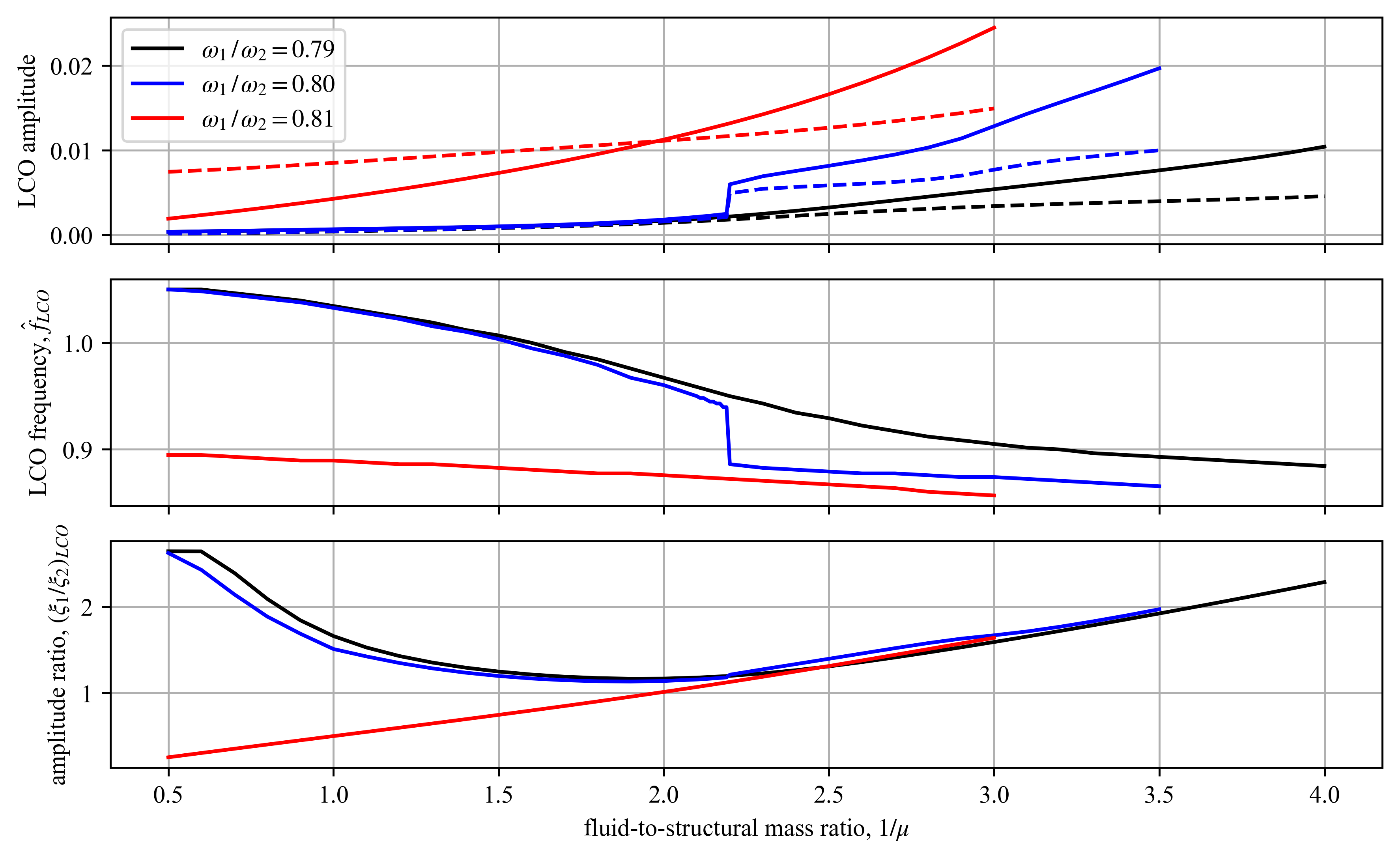}
    \caption{LCO amplitude, frequency, and amplitude ratio as a function of fluid-to-structural mass ratio for various mode 1-to-2 natural frequency ratios ($\omega_1/\omega_2$) with $\hat{f}_2 = 0.90$ and $\zeta_1=\zeta_2=0.00$.}
    \label{fig:IFASD_coupled_modeA}
\end{figure}

Figure~\ref{fig:IFASD_coupled_modeA} shows the response amplitude, frequency and amplitude ratio for $\hat{f}_2 = 0.90$, with no structural damping, and very modest changes in $\omega_1\, / \, \omega_2$. A reasonably large modal velocity perturbation is applied as an initial condition to ensure that any potential subcritical instabilities are identified. Two distinct behaviors can be observed. These are described as follows:

\begin{enumerate}

\item \textbf{Supercritical}: For natural frequency ratios $\omega_1\, / \, \omega_2 \leq 0.79$, the system is stable at low values of $1/\mu$ (oscillating at a low amplitude at the buffet frequency), with a moderate dominance of mode 1. As $1/\mu$ increases, the response amplitude grows slowly, the LCO frequency reduces (towards $\hat{f}_2$), and the amplitude ratio drops towards unity (mode 2 becomes increasingly prominent). Eventually, as the $1/\mu$ becomes large and the LCO frequency approaches the mode 2 natural frequency, faster growth of the LCO is observed through mode 1, although the LCO growth (as a function of $1/\mu$) and amplitude are still low compared to the subcritical regime described below. The LCO amplitude ratio for this response regime is $\mathcal{O}(1)$. 

\item \textbf{Subcritical}: For mode 1-to-2 natural frequency ratios $\omega_1\, / \, \omega_2 \geq 0.81$, the system is unstable down to the limit of zero fluid-to-structural mass ratio, $1/\mu = 0$. This instability coexists with the low-amplitude buffet-frequency branch described above, so the response is subcritical and a substantial (but well within physical bounds) perturbation is required to induce the instability. At low values of $1/\mu$ the LCO is mode 2 dominated (but not overwhelmingly so), both in frequency and amplitude. As $1/\mu$ increases, the mode 1 LCO amplitude grows much more rapidly than that of mode 2, and while the LCO amplitude of both modes becomes very large. As mode 1 dominance increases, the LCO frequency drops below the mode 2 natural frequency, although remains closer to the mode 2 natural frequency than the mode 1 natural frequency. The LCO amplitude ratio for this response regime is predominantly $\mathcal{O}(1)$ and varies linearly with $1/\mu$.

\item \textbf{Transitional}: Most interestingly and importantly, at the boundary between the two response regimes, it is possible to have a system which exhibits both behaviors, as can be observed at $\omega_1\, / \, \omega_2 = 0.80$. This system demonstrates the behavior of the \textbf{Supercritical} system at low fluid-to-structural mass ratios, and abruptly transitions to that of the \textbf{Subcritical} system as the mass ratio increases. 

\end{enumerate}

\clearpage

\begin{figure}[!h]
    \centering
    \includegraphics[width=1.0\textwidth]{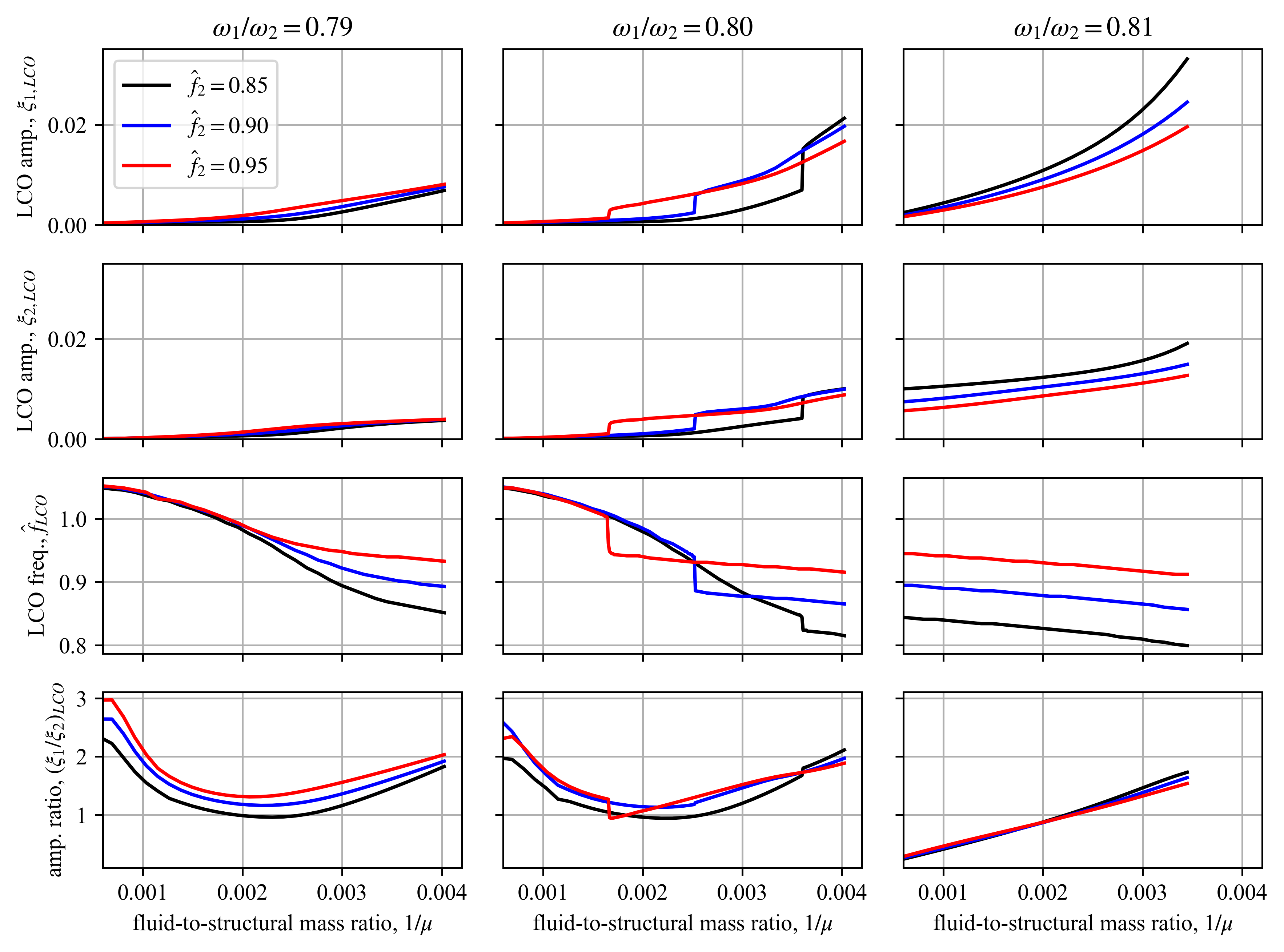}
    \caption{LCO amplitude, frequency, and amplitude ratio as a function of fluid-to-structural mass ratio for various mode 2 natural frequencies ($\hat{f}_2$) and mode 1-to-2 natural frequency ratios ($\omega_1/\omega_2$) with $\zeta_1=\zeta_2=0.00$.}
    \label{fig:IFASD_coupled_modeB}
\end{figure}

Figure~\ref{fig:IFASD_coupled_modeB} shows how the three response regimes vary as a function of the mode 2 natural frequency ratio, $\hat{f}_2$, and the mode 1-to-2 natural frequency ratio, $\omega_1/\omega_2$. It is immediately clear that the character of the regimes is entirely dependent on $\omega_1/\omega_2$, rather than the mode 2 natural frequency. For instance, the \textbf{supercritical} regime at $\omega_1/\omega_2 = 0.79$ is barely influenced by quite substantial changes in $\hat{f}_2$. What this means, physically, is that in this regime, the response is less influenced by the vicinity of the structural natural frequency to the buffet frequency. At $\omega_1/\omega_2 = 0.80$, although the general behavior of the \textbf{transitional} response regime is also insensitive to $\hat{f}_2$, the transition point changes substantially, meaning it is strongly dependent on the nearness of the mode 2 natural frequency to the buffet frequency. This is behavior is generally consistent with that of single-degree-of-freedom subcritical aeroelastic instabilities in buffet: the instability is easier to trigger and less aggressive when the detuning is smaller (buffet and structural natural frequencies are closer). For the \textbf{subcritical} response, both $\xi_{1,LCO}$ and $\xi_{2,LCO}$ increase monotonically as $\hat{f}_2$ is reduced from 0.95 to 0.85, and the LCO frequency, $\hat{f}_{LCO}$, is offset accordingly. Despite this, the amplitude ratio, $(\xi_1/\xi_2)_{LCO}$, is largely insensitive to $\hat{f}_2$, with the three curves collapsing onto an approximately common trend. This indicates that, in this regime, $\hat{f}_2$ governs the overall magnitude and frequency of the LCO, but not the relative participation of the two modes, which appears to be primarily governed by $\omega_1/\omega_2$.

At this point evidence for and against coupled-mode instabilities has been presented. The evidence to support the presence of a coupled-mode instability is (i) the modal LCO amplitude ratio being $\mathcal{O}(1)$, and (ii) the presence of a \textbf{supercritical} regime and a stable response without structural damping. The evidence against the presence of a coupled-mode instability is (i) the LCO frequency being very close to the mode 2 natural frequency, and (ii) the presence of a \textbf{subcritical} regime and, in the absence of structural damping, an unstable response down to $1/\mu = 0$. 

\begin{figure}[!h]
    \centering
    \includegraphics[width=1.0\textwidth]{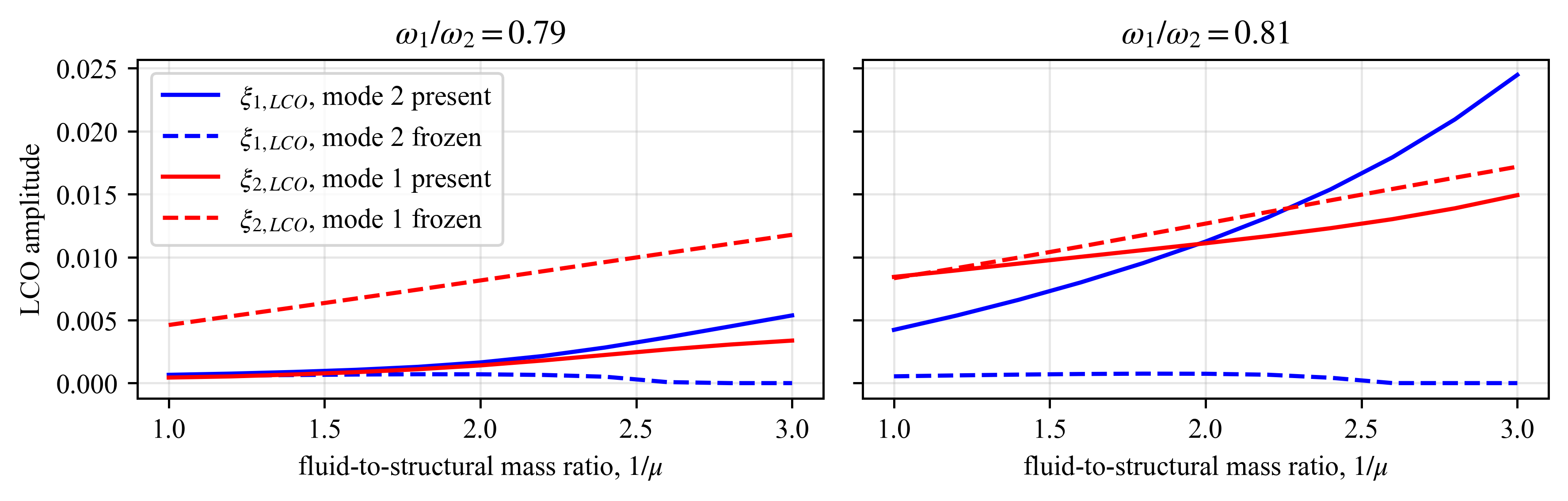}
    \caption{LCO amplitude with the other mode frozen for the supercritical and subcritical regimes with $\hat{f}_2 = 0.90$ and $\zeta_1=\zeta_2=0.00$.}
    \label{fig:IFASD_coupled_modeD}
\end{figure}

Although there are several approaches for further investigations, one that is most attractive is to perform computations for each mode in isolation (with the other mode frozen), as presented in Fig.~\ref{fig:IFASD_coupled_modeD}. If the instability persists through a single mode, then it can be diagnosed as a single-degree-of-freedom instability. This is repeated for both response regimes. For the \textbf{subcritical} regime ($\omega_1/\omega_2 = 0.81$), the mode 2 instability persists almost unchanged when mode 1 frozen. Conversely, when mode 2 is frozen, mode 1 becomes stable. From this, the subcritical instability can be very clearly diagnosed as single-degree-of-freedom, and the modal LCO amplitude ratio $\mathcal{O}(1)$ emerges from the unstable mode 2 pulling mode 1. For the \textbf{supercritical} regime ($\omega_1/\omega_2 = 0.79$) the picture is not as simple. When mode 1 is frozen, the mode 2 response regime shifts to a conventional single-degree-of-freedom subcritical instability. Conversely, when mode 2 is frozen, mode 1 becomes stable. This suggests that the presence of an elastic mode 1 acts as an energy sink, changing the character of the response from aggressive and subcritical to supercritical. This is supported by the amplitude comparison in Fig.~\ref{fig:IFASD_coupled_modeD}: mode 2 remains substantially below its isolated response across the entire swept range of $1/\mu$, and experiences only smooth growth as $1/\mu$ increases. Mode 1 also grows steadily with $1/\mu$ despite being stable in isolation. 

\begin{figure}[!h]
    \centering
    \includegraphics[width=0.625\textwidth]{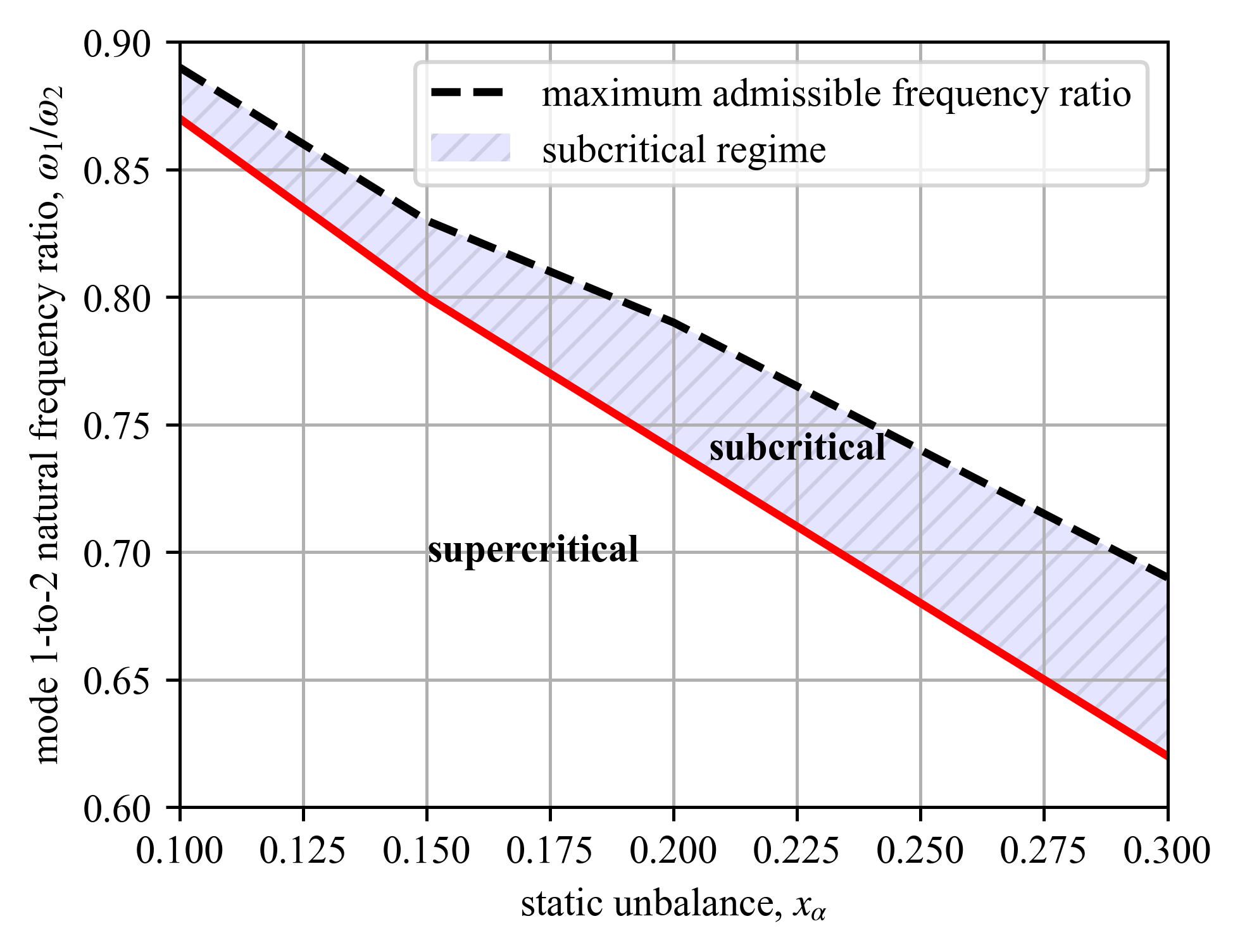}
    \caption{Supercritical and subcritical regions as a function of fluid-to-structural mass ratio for various mode 2 natural frequencies ($\hat{f}_2$) and mode 1-to-2 natural frequency ratios ($\omega_1/\omega_2$) with $\hat{f}_2 = 0.90$ with $\zeta_1=\zeta_2=0.00$.}
    \label{fig:IFASD_coupled_modeE}
\end{figure}

Figure~\ref{fig:IFASD_coupled_modeE} summarizes the influence of static unbalance on the boundary between the two response regimes. The dashed line denotes the maximum admissible mode 1-to-2 natural frequency ratio, above which the inverse structural eigenvalue problem does not admit real, positive stiffness values. As $x_\alpha$ increases, both the supercritical-to-subcritical boundary and the maximum admissible frequency ratio shift toward lower values of $\omega_1/\omega_2$, while the range of frequency ratios associated with the subcritical response broadens. Thus, greater static unbalance allows the subcritical, mode-2-driven branch to occur when the structural frequencies are more widely separated, whereas lower values of $\omega_1/\omega_2$ retain the supercritical response in which interaction with mode 1 suppresses the isolated mode 2 instability. Above the dashed boundary, the resulting amplitudes or frequencies exceed the validated operating range of the ROM.

To summarize this section, coupled-mode instabilities, in a classical flutter sense, have not been identified. However, response regimes where coupling between modes exists have been. Specifically, a supercritical regime has been identified which is characterized by the stabilizing influence of a heave-dominated mode on a pitch-dominated mode. Importantly, this regime can also abruptly transition to an aggressive subcritical regime. The other mode coupling mechanism identified is a single-degree-of-freedom instability of a pitch-dominated mode exciting a stable heave dominated mode when their natural frequencies are close to each other.
\clearpage

\section{Conclusion}

A physics-guided multi-input Neural DE has been developed for nonlinear aeroelastic systems driven by self-excited flows. By combining a Rayleigh oscillator, finite-memory Volterra terms, and a compact neural correction, the model learns multi-mode aerodynamic dynamics from a single prescribed-motion CFD simulation and accurately predicts stability, lock-in, and LCO amplitude and frequency. For transonic buffet, the framework reproduced both large-amplitude pitching responses and mixed heave--pitch dynamics over a broad parameter range. The results show that buffet aeroelasticity is primarily governed by single-mode instabilities, even when the physical motion contains strong heave--pitch coupling. Classical coupled-mode flutter was not identified; instead, modal interaction can suppress an otherwise subcritical instability, produce a smooth supercritical response, or allow an unstable pitch-dominated mode to excite a stable heave-dominated mode. These findings demonstrate that the proposed ROM provides both an efficient predictive tool and a practical means of exposing nonlinear modal interactions that would be prohibitively expensive to investigate using CFD alone.

\section*{Acknowledgments}
This work is supported by the Asian Office of Aerospace Research and Development (AOARD) and Air Force Office of Scientific Research (AFOSR) for project FA2386-24-1-4044: Data-Driven Reduced Order Modelling and Preliminary Experimentation for Combined Transonic Buffet and Freeplay Induced Limit Cycle Oscillations and the partial financial support provided by Australian Defence Science and Technology Group (DSTG). The contributions of our partners on the project, Prof. Francesco Toffol and Prof. Sergio Ricci from Politecnico di Milano, and Prof. Natsuki Tsushima from Kyushu University have been instrumental to this work. The generous resources and support of Dale Osborne and Dr. Robert Shen from RMIT RACE Cloud Supercomputing Hub, and the resources provided by ANSYS and the support of Dr Valerio Viti and Dr Luke Munholand, are also all greatly appreciated.

\bibliographystyle{elsarticle-num} 
\bibliography{gensys_doi}
        

\end{document}